\documentclass[authoryear,final,3p,times]{elsarticle}
\usepackage{amssymb}
\usepackage{amsthm}
\usepackage{amsmath}
\usepackage{bm}
\usepackage{array}
\usepackage{ulem}
\usepackage{longtable}
\usepackage{CJK}
\usepackage{setspace}
\usepackage{multirow}
\usepackage{siunitx}
\usepackage{threeparttable}
\usepackage{verbatim}
\usepackage{rotating}
\usepackage{xcolor}
\usepackage{booktabs}
\usepackage[justification=centering]{caption}
\usepackage{lineno}
\usepackage{setspace}
\usepackage{lscape}
\usepackage{makecell}
\usepackage{threeparttable}
\usepackage{float}
\usepackage{graphicx}
\usepackage{lineno}
\allowdisplaybreaks[4]

\begin{document}
%\linenumbers
%\pagewiselinenumbers
%\switchlinenumbers

\begin{spacing}{1.0}
\begin{frontmatter}

%\journal{Chemical Engineering Science}

\title{ Hydrodynamics of polydisperse gas-solid flows: Kinetic theory and multifluid simulation}

\author[label1,label2]{Bidan Zhao\corref{cor1}}
\ead{bdzhao@ipe.ac.cn}
\author[label1,label3]{Kun Shi}
\author[label1,label2]{Mingming He}
\author[label1,label2]{Junwu Wang\corref{cor1}}
\cortext[cor1]{Corresponding author}
\ead{jwwang@ipe.ac.cn}

\address[label1]{State Key Laboratory of Multiphase Complex Systems, Institute of Process Engineering, Chinese Academy of Sciences, P. O. Box 353, Beijing 100190, P. R. China}
\address[label2]{School of Chemical Engineering, University of Chinese Academy of Sciences, Beijing, 100049, P. R. China}
\address[label3]{School of Chemical Engineering, Shenyang University of Chemical Technology, Shenyang 110142, P. R. China}
%\address[label4]{Innovation Academy for Green Manufacture, Chinese Academy of Sciences, P. O. Box 353, Beijing 100190, P. R. China}

\begin{abstract}
Polydisperse gas-solid flows, which is notoriously difficult to model due to the complex gas-particle and particle-particle interactions, are widely encountered in industry. In this article, a refined kinetic theory for polydisperse flow is developed, which features single-parameter Chapman-Enskog expansion (the Knudsen number) and exact calculation of the integrations related to pair distribution function of particle velocity without any mathematical approximations. The Navier-Stokes order constitutive relations for multifluid modeling of polydisperse gas-solid flow are then obtained analytically, including the solid stress tensor, the solid-solid drag force, the granular heat flux and the energy dissipation rate. Finally, the model is preliminarily validated by comparing to the discrete element simulation data of one-dimensional granular shear flow and
by showing that the hydrodynamic characteristics of gas-solid flows in a bubbling fluidized bed containing bidisperse particles can be successfully predicted.

\end{abstract}
\end{frontmatter}
\section{Introduction}
Particles of different densities/sizes are widely encountered in industrial reactors and in nature \citep{hrenya2011kinetic}. This characteristic
is called polydispersity. Polydisperse particles have a significant influence on the hydrodynamics of
complex gas-solid flows \citep{grace1991influence}, such as the mixing and segregation phenomena of different particle species \citep{sundaresan2001some,rosato1987brazil}. In particular, understanding of the polydisperse effect facilitates the design and optimization of fluidized bed reactors \citep{sun1990effect,grace1991influence}. For example, the complete mixing
of drug and binder particles is essential for the control of drug delivery in the
tablet formation process, and the segregation of large and small particles facilitates mineral sorting \citep{zhang2020cfd}.

Computer simulation has become a favourite approach to study the hydrodynamics of polydisperse flows in recent decades \citep{Beetstra2007Drag,he2009gas,berger2014challenges,lan2020long,lan2021scale}. Among them, continuum models achieve a balance between simulation accuracy and efficiency for simulating industrial-scale reactors \citep{wang2020continuum}, where the constitutive relations of particulate phase stresses and solid-solid drag forces are typically closed using polydisperse kinetic theory of granular flow. The kinetic-theory-based continuum models of polydisperse gas-solid flows can be classified into the mixture model \citep{jenkins1989kinetic,willits1999kinetic,garzo2019navier,V2021Comment,van2009developmentI,van2009developmentII} and species model \citep{lathouwers2000modeling,mathiesen2000predictions,huilin2003hydrodynamic,Iddir2004modeling,chao2011derivation,solsvik2021kinetica, solsvik2021kineticb, zhao2021kinetic,shi2022critical}. The mixture model comprises a mass balance equation for each solid species, as well as a momentum conservation equation and a granular temperature equation for the solid mixture. When the solid species number is $N$, its total number of governing equations is $N+4$. As a result, the volume fraction of each species, the average velocity and the granular temperature of the solid mixture can be determined.
It is widely agreed that non-equilibrium factors, as expressed by the perturbed particle velocity distribution function (PVDF), contribute to the differences in velocity and granular temperature between different solid species. These distinctions can be algebraically deduced rather than incurring additional costs from improving computational calculations with partial differential equations \citep{garzo2019navier}. Therefore, the primary advantage of the mixture model is cost reduction, while its disadvantage stems from the assumption that non-equilibrium effects are the sole cause of discrepancies in velocity and granular temperature, and the simplified derivation of the perturbation function \citep{van2009developmentI,van2009developmentII}. As software and hardware capabilities improve, multifluid models, where species kinetic theory is used, are frequently used to alleviate those limitations in the simulation of polydisperse systems with the price of higher computational cost. Species models incorporate the mass balance equation, momentum conservation equation, and granular temperature equation for each species, resulting in a total of $5N$ conservation equations.

The accuracy of multifluid model critically depends on the gas-solid interactions and the particulate constitutive relations \citep{shu2014multifluid,zhou2015cfd,qin2019emms}, where polydisperse kinetic theory of granular flow (KTGF) is a widely employed technique for theoretically closing particulate constitutive relations such as the solid stress tensors and the solid-solid drag forces \citep{hrenya2011kinetic}. It is widely acknowledged that the particle velocity distribution function plays a critical role in polydisperse KTGF, since the kinetic part related to microscale fluctuation and the collisional part related to particle-particle interactions in these constitutive relations can be expressed as the integrations of particle velocity related properties. Another deeper reason is that a PVDF should be consistent with the physics of the studied system, specifically, the Maxwellian distribution corresponds to the equilibrium state \citep{chao2011derivation}, the homogeneous cooling distribution correlates to the non-equilibrium but homogeneous state \citep{garzo2007enskog,garzo2019navier}, and the bimodal distribution pertains to the strong non-equilibrium state of heterogeneous gas-solid flows \citep{zhao2018unification,zhao2021multiscale}. The Chapman-Enskog expansion method is then normally used to determine the PVDF considering the perturbation effect \citep{chapman1970mathematical,Iddir2004modeling,iddir2005analysis,iddir2005modeling,zhao2021kinetic}.
There are in general two different approaches to achieve the Chapman-Enskog expansion, one is expanded using the Knudsen number $Kn$ as the sole parameter to study the combined impact of non-equilibrium and inelastic particle-particle collision \citep{Iddir2004modeling,iddir2005analysis,iddir2005modeling}, the other use dual-parameter expansion ($Kn$ and $\epsilon\equiv(1-e_{ij}^2)$ that quantifies the energy dissipation due to inelastic collisions) to distinguish the non-equilibrium effect from the inelastic effect \citep{sela1998hydrodynamic,serero2006hydrodynamics,zhao2021kinetic,shi2022critical}. As systematically reviewed in a previous study \citep{shi2022critical}, those approaches may result in significantly different constitutive relations. On the basis of comparison between the results using the single-parameter or dual-parameter expansion and those obtained from the discrete element simulations of binary granular flow, it was concluded that the constitutive relation of \cite{iddir2005modeling} that is obtained using single-parameter expansion method is best \citep{shi2022critical}.
However, the work of \cite{iddir2005modeling} contains mathematical and/or physical approximations during the derivation of constitutive relations, {{such as utilizing truncated Taylor series expansion for complex collision integrations, disregarding the contribution to the perturbation function due to the interaction of particles from different species and neglecting the collisional contribution of perturbation function for the constitutive relations at Navier-Stokes order}}.

In the present work, the polydisperse kinetic theory of granular flow of \cite{iddir2005modeling} is refined in the sense that the complex collision integrations are calculated completely and exactly without any mathematical approximations by using the mathematical technique we developed previously \citep{zhao2020note,zhao2021kinetic}, whereas the perturbation function still neglects the contribution of the interaction of particles from different species as in the study of \cite{iddir2005modeling}. Moreover, there are minor differences related to the effect of particle
inelasticity in the specific forms of the perturbation functions as compared to \cite{iddir2005modeling}.
{The theoretically predicted constitutive relations are then preliminarily validated using the discrete element results of one-dimensional, shear granular flow of \cite{galvin2007hydrodynamic}, where the results predicted by \cite{iddir2005modeling} are also presented. Finally, present polydisperse KTGF is coupled with multifluid model to analyze the hydrodynamics of complex gas-solid flow in a bubbling fluidized bed.}

\section{Multifluid model}
The derivation of the hydrodynamic governing equations and the constitutive relations from the granular kinetic theory is performed as follows: (i) The Boltzmann equation with an appropriate source term for particle collisions controls the spatio-temporal evolution of the PVDF; (ii) The motion properties of individual particles, such as mass, momentum, and fluctuate energy, are multiplied by both sides of the Boltzmann equation. This results in the derivation of the Maxwellian equation, a macroscopic transport equation obtained by averaging the motion states of all particles through the integration of microscopic particle velocity; (iii) After defining the particle properties, we can derive the mass balance equation, {{the}} momentum conversation equation, and {{the}} granular temperature equation in sequence. This process enables us to establish the constitutive relations, including the solid stress tensor, the solid-solid drag force, the heat flux, and the energy dissipation rate. Technical details of this process are similar to our previous work \citep{zhao2021kinetic}, therefore, they are not given here, instead only the final form of the hydrodynamic governing equations and the constitutive relations are directly reported below.

\subsection{Governing equations}
The governing equations for solid species $i$ in the multifluid model include:\\
(i) mass balance equation:
\begin{equation}
\begin{split}
\frac{\partial}{\partial t}(\varepsilon_i\rho_i)+\nabla\cdot(\varepsilon_i\rho_i{\bf{u}}_i)=0,
\end{split}
\label{a1}
\end{equation}
(ii) momentum conversation equation:
\begin{equation}
\begin{split}
\frac{\partial}{\partial t}(\varepsilon_i\rho_i{\bf{u}}_i) + \nabla\cdot(\varepsilon_i\rho_i{\bf{u}}_i{\bf{u}}_i)
=\varepsilon_i\rho_i<{\bf{F}}_i>-\nabla\cdot{\bf{p}}_i
+\sum_{j=1}^N{\bm{\mathcal{F}}}_{D,ij},
\end{split}
\label{a2}
\end{equation}
(iii) granular temperature equation:
\begin{equation}
\begin{split}
\frac{3}{2}\bigg[\frac{\partial}{\partial t}
(n_iT_i)+\nabla\cdot(n_iT_i{\bf{u}}_i)\bigg]
=-{\bf{p}}_i:\nabla{\bf{u}}_i+\nabla\cdot{\bf{q}}_i
+{\mathcal{N}}_{d,i}-\sum_{j=1}^N{\bf{u}}_i\cdot{\bm{\mathcal{F}}}_{D,ij}
+m_in_i<{\bf{F}}_i\cdot{\bf{C}}_i>,
\end{split}
\label{a3}
\end{equation}
where the volume fraction, the density, the averaged velocity, the granular temperature of species $i$ are expressed by $\varepsilon_i$, $\rho_i$, ${\bf{u}}_i$, $T_i$. The two parts associated with the external force {{${\bf{F}}_i$}} are:
\begin{equation}
\begin{split}
\varepsilon_i\rho_i<{\bf{F}}_i>=\varepsilon_i\rho_i\dot{\bf{G}}+\beta_{A,i}({\bf{u}}_g-{\bf{u}}_i)-\varepsilon_i\nabla P_g,
\end{split}
\label{a4}
\end{equation}
\begin{equation}
\begin{split}
m_in_i<{\bf{F}}_i\cdot{\bf{C}}_i>=\beta_{A,i}<{\bf{C}}_g\cdot{\bf{C}}_i>-\beta_{A,i}\frac{3T_i}{m_i},
\end{split}
\label{a5}
\end{equation}
where $\dot{\bf{G}}$ is the gravitational acceleration, $\beta_{A,i}$ is the gas-solid drag coefficient between gas and species $i$, ${\bf{u}}_g$ is the macroscopic velocity of gas, $P_g$ is the gas pressure, $m_i$ is the particle mass of species $i$, $n_i$ is the number density of species $i$ ($m_in_i=\varepsilon_i\rho_i$), ${\bf{C}}_g$ is the fluctuation velocity of gas molecule, ${\bf{C}}_i\equiv{\bf{c}}_i-{\bf{u}}_i$ is the fluctuation velocity of particle and $<\cdot>$ is the averaged operator which is defined as $<\varphi>\equiv\frac{\int\varphi f_id{\bf{c}}_i}{\int f_id{\bf{c}}_i}$. In present study, $\beta_{A,i}<{\bf{C}}_g\cdot{\bf{C}}_i>$ has been neglected following a previous study \citep{gidaspow1994multiphase}.

We focus on how the interaction between particles and the stochastic velocity fluctuation of particles contribute to the solid stress tensor ${\bf{p}}_i$, the solid-solid drag force ${\bm{\mathcal{F}}}_{D,ij}$, the heat flux ${\bf{q}}_i$ and the energy dissipation rate ${\mathcal{N}}_{d,i}$. They can be obtained {{on the basis of the PVDF $f_i$ }}as follows:
\begin{equation}
\begin{split}
{\bf{p}}_i
=&m_i\int{\bf{C}}_i{\bf{C}}_if_id{\bf{c}}\\
+&\sum_j^N\bigg\{\frac{d_{ij}^3}{2}g_{ij}m_i\int\int\int\limits_{({\bf{g}}\cdot{\bf{k}})>0} ({\bf{c}}_i^{'}-{\bf{c}}_i)f_if_j{\bf{k}}({\bf{k}}\cdot{\bf{g}})d{\bf{k}}d{\bf{c}}_id{\bf{c}}_j\\
+&\frac{d_{ij}^4}{4}g_{ij}m_i\int\int\int\limits_{({\bf{g}}\cdot{\bf{k}})>0}
({\bf{c}}_i^{'}-{\bf{c}}_i)f_if_j\nabla\ln\frac{f_i}{f_j}\cdot({\bf{k}}\cdot{\bf{g}}){\bf{kk}}d{\bf{k}}d{\bf{c}}_id{\bf{c}}_j\bigg\},
\end{split}
\label{a6}
\end{equation}
\begin{equation}
\begin{split}
{\bm{\mathcal{F}}}_{D,ij}
=&d_{ij}^2g_{ij}m_i\int\int\int\limits_{({\bf{g}}\cdot{\bf{k}})>0}
({\bf{c}}_i^{'}-{\bf{c}}_i)f_if_j({\bf{g}}\cdot{\bf{k}})d{\bf{k}}d{\bf{c}}_id{\bf{c}}_j\\
+&\frac{d_{ij}^3}{2}g_{ij}m_i\int\int\int\limits_{({\bf{g}}\cdot{\bf{k}})>0}
({\bf{c}}_i^{'}-{\bf{c}}_i)f_if_j\nabla\ln\frac{f_i}{f_j}\cdot{\bf{k}}({\bf{g}}\cdot{\bf{k}})
d{\bf{k}}d{\bf{c}}_id{\bf{c}}_j,
\end{split}
\label{a7}
\end{equation}
\begin{equation}
\begin{split}
{\bf{q}}_i
=&\frac{1}{2}m_i\int{\bf{C}}_i^2{\bf{C}}_if_id{\bf{c}}\\
+&\sum_j^N\bigg\{\frac{d_{ij}^3}{4}g_{ij}m_i\int\int\int\limits_{({\bf{g}}\cdot{\bf{k}})>0} ({\bf{C}}_i^{'2}-{\bf{C}}_i^2)f_if_j{\bf{k}}({\bf{k}}\cdot{\bf{g}})d{\bf{k}}d{\bf{c}}_id{\bf{c}}_j\\
+&\frac{d_{ij}^4}{8}g_{ij}m_i\int\int\int\limits_{({\bf{g}}\cdot{\bf{k}})>0}
({\bf{C}}_i^{'2}-{\bf{C}}_i^2)f_if_j\nabla\ln\frac{f_i}{f_j}\cdot({\bf{k}}\cdot{\bf{g}}){\bf{kk}}d{\bf{k}}d{\bf{c}}_id{\bf{c}}_j\bigg\},
\end{split}
\label{a8}
\end{equation}
and
\begin{equation}
\begin{split}
\mathcal{N}_{d,i}
=&\sum_j^N\bigg\{d_{ij}^2g_{ij}m_i\int\int\int\limits_{({\bf{g}}\cdot{\bf{k}})>0}
({\bf{c}}_i^{'2}-{\bf{c}}_i^2)f_if_j({\bf{g}}\cdot{\bf{k}})d{\bf{k}}d{\bf{c}}_id{\bf{c}}_j\\
+&\frac{d_{ij}^3}{2}g_{ij}m_i\int\int\int\limits_{({\bf{g}}\cdot{\bf{k}})>0}
({\bf{c}}_i^{'2}-{\bf{c}}_i^2)f_if_j\nabla\ln\frac{f_i}{f_j}\cdot{\bf{k}}({\bf{g}}\cdot{\bf{k}})
d{\bf{k}}d{\bf{c}}_id{\bf{c}}_j\bigg\}.
\end{split}
\label{a9}
\end{equation}
The flux terms including the solid stress tensor (also called as the momentum flux) and the heat flux can be split into the kinetic contribution related to the fluctuation velocity of particles and the collisional contribution related to particle-particle contact interaction. On the other hand, the source terms including the solid-solid drag force and the energy dissipation rate are only related to the inelastic particle-particle collisions. As shown in the fundamental forms of Eq.(\ref{a6})-Eq.(\ref{a9}), these constitutive relations of polydisperse flow are complex because it is difficult to analytically treat the integrals associated with the particle-particle interaction between different species, which needs to consider the differences in the average velocities and the granular temperatures. Thus, in most of the previous works of species KTGF for polydisperse systems \citep{chao2011derivation,huilin2001kinetic,mathiesen2000predictions}, only the Maxwellian PVDF is substituted into these expressions without considering the contributions of non-equilibrium effect that are mathematically represented via the perturbation functions. Even after this simplification, the numerical integration technique \citep{chao2011derivation} and the approximation of integrants by Taylor series expansion \citep{huilin2001kinetic,mathiesen2000predictions,Iddir2004modeling} were still used to calculate the complex integrals. According to a previous work \citep{shi2022critical}, the prediction ability of the constitutive relations that are derived using the Maxwellian PVDF is inferior to those obtained from the PVDF with perturbation functions obtained through the Chapman-Enskog method, and the constitutive relations of \cite{iddir2005modeling} that is obtained using single-parameter Chapman-Enskog method is best in term of the comparison to the discrete element simulation results of binary shear granular flow.
However, their constitutive relations still involves a variety of approximations \citep{Iddir2004modeling,iddir2005analysis}: (i) the effects of the interaction between different types of particles on the perturbation function are completely neglected; (ii) the Navier-Stokes order constitutive relations related to the interaction between different types of particles only cover the contribution of Maxwellian PVDF, and the contribution of non-equilibrium effects via the relevant perturbation functions is not considered; and finally, (iii) the Taylor series expansion approximates the complex interactions, but with the increasing differences in the interphase slip velocity and the granular temperatures, the neglected terms are significant and directly result in larger calculation errors \citep{zhao2020note}.

Here, we refine the work of \cite{iddir2005modeling} via removing the approximation of points (ii) and (iii) just mentioned to obtain mathematically rigorous and analytical constitutive relations, which is realized by using the mathematical technique that we developed previously \citep{zhao2020note,zhao2021kinetic}. However, the removal of approximation (i) leaves for future study, since we do not know how to fix it yet.

\subsection{Single-parameter Chapman-Enskog method}
The polydisperse system at the state away from the equilibrium is quantified by the non-local gradients of macroscopic hydrodynamic variables ( $\nabla \varepsilon_i$, $\nabla{\bf{u}}_i$, $\nabla T_i$) and the differences in volume fractions, average velocities and granular temperatures of different species in KTGF. The single-parameter Chapman-Enskog method unites these factors in the perturbation function $\Phi_i^{(K)}$. Based on the Maxwellian PVDF ($f_i^{(0)}$), the particle velocity distribution function of species $i$ in polydisperse system away from the equilibrium takes the following form \citep{Iddir2004modeling,iddir2005analysis,iddir2005modeling,chapman1970mathematical}:
\begin{equation}
\begin{split}
&f_i=f_i^{(0)}(1+\Phi_i^{(K)}+\cdots),
\end{split}
\label{a10}
\end{equation}
where
\begin{equation}
\begin{split}
&f_i^{(0)}=n_i(\frac{m_i}{2\pi T_i})^{\frac{3}{2}}
\exp(-\frac{m_i({\bf{c}}_i-{\bf{u}}_i)^2}{2T_i}),
\end{split}
\label{a10-1}
\end{equation}
where the parameter $K$ indicates that the extent of the system's deviation from the equilibrium state is only associated with the Knunsen number. The perturbation function $\Phi_i^{(K)}$ incorporates all first-order effects of  gradients or differences (i.e., Navier-Stokes order), with its form being determined by the Boltzmann equation:
\begin{equation}
\begin{split}
\frac{\partial f_i }{\partial t}+{\bf{c}}_i\cdot\frac{\partial f}{\partial {\bf{r}}}+\frac{\partial }{{\partial {\bf{c}}}_i}\cdot({\bf{F}}_if_i)=\bigg(\frac{\partial f_i}{\partial t}\bigg)_{coll}.
\end{split}
\label{a11}
\end{equation}
Specifically, upon substituting the {{PVDF}} (\ref{a10}) into the Boltzmann equation, the terms on both sides have been retained up to the first order of the parameter $K$ are:
\begin{equation}
\begin{split}
\sum\limits _{j=1}^N n_in_j\mathcal{I}(\Phi_i^{(K)},\Phi_j^{(K)})
=\sum\limits_{j=1}^N J_1(f_i^{(0)},f_j^{(0)})-{\mathcal{D}}_kf_i^{(0)},\\
\end{split}
\label{a12}
\end{equation}
where the terms related to the changes in distribution function due to transient, convective, and external-force are denoted as ${\mathcal{D}}_Kf_i^{(0)}$, and the linearized Boltzmann operator is defined as
\begin{equation}
\begin{split}
&\mathcal{I}(\Phi_i^{(K)},\Phi_j^{(K)})
\equiv\frac{g_{ij}}{n_i n_j}\int\int f_i^{(0)}f_j^{(0)}(\Phi_i^{(K)}+\Phi_j^{(K)}-\Phi_i^{(K)'}-\Phi_j^{(K)'})
d_{ij}^2({\bf{g}}\cdot{\bf{k}})d{\bf{k}}d{\bf{c}}_j,\\
\end{split}
\label{a13}
\end{equation}
and based on the Maxwellian PVDF, the remaining part related to particle collisions can be written as follows
\begin{equation}
\begin{split}
J_1(f_i^{(0)},f_j^{(0)})
\equiv&d_{ij}\int\int  f_i^{(0)}f_j^{(0)}
[{\bf{k}}g_{ij}\cdot\ln(f_j^{(0)'}f_j^{(0)'})+{\bf{k}}\cdot \nabla g_{ij}
]d_{ij}^2({\bf{g}}\cdot{\bf{k}})d{\bf{k}}d{\bf{c}}_j.\\
\end{split}
\label{a14}
\end{equation}
We have been attempting to determine the perturbation function from the complete governing equation (\ref{a12}) for an extended period. Tragically, no methods  was found to handle the complex form analytically. As a result, this study opts for the simplified governing equation for the perturbation function. Therefore, the simplified governing equation of perturbation function is selected as follows in this study \citep{chapman1970mathematical}:
\begin{equation}
\begin{split}
n_i^2\mathcal{I}(\Phi_i^{(K)},\Phi_i^{(K)})
= J_1(f_i^{(0)},f_i^{(0)})-{\mathcal{D}}_kf_i^{(0)},\\
\end{split}
\label{a15}
\end{equation}
where the effects of particle collisions between different species have been neglected, accurate solution of the perturbation function
can then be achieved by inducing the Sonnie polynomials \citep{chapman1970mathematical}:
\begin{equation}
\begin{split}
\Phi_i^{(K)}
\equiv&-A_i\bigg\{B_i(\frac{{\bf{C}}_i^2}{2\theta_i}-\frac{5}{2}){\bf{C}}_i\cdot\nabla\ln
\theta_i
+D_i({\bf{C}}_i{\bf{C}}_i-\frac{1}{3}{\bf{C}}_i^2{\bf{I}}):\nabla{\bf{u}}_i\bigg\},
\end{split}
\label{a16}
\end{equation}
where
\begin{equation}
\begin{split}
&A_i=\frac{1}{n_ig_{ii}}\frac{15}{32d_i^2}(\theta_i\pi)^{-\frac{1}{2}}\\
&B_i=(1+\frac{2}{5}n_i\pi d_i^3g_{ii})\\
&D_i={{\frac{2}{3\theta_i}(1+\frac{4}{15}n_i\pi d_i^3g_{ii})}}.
\end{split}
\label{a17}
\end{equation}
It's noted that there are minor differences in these associated coefficients of the perturbation function between present study and the work of \cite{iddir2005modeling}, specifically, $B_{i,Iddir}\equiv\bigg[1+\frac{1}{10}(1+e)^2n_i\pi d_i^3 g_{ii}\bigg]$ and $D_{i,Iddir}\equiv{{\frac{2}{3\theta_i}\bigg[1+\frac{2}{15}(1+e)n_i\pi d_i^3g_{ii}\bigg]}}$.
Their perturbation function is also not an analytical and accurate solution of Eq.(\ref{a12}), although it has taken the contribution from the inelastic particle collision into account by modifying the basic form of the perturbation function proposed by \cite{chapman1970mathematical}.

Based on the relationships between the macroscopic hydrodynamic variables and the Maxwellian PVDF, we have established the definitions of these variables:
\begin{equation}
\begin{split}
n_i=\int f_i^{(0)}d{\bf{c}}_i, \quad n_{i}{\bf{u}}_i=\int{\bf{c}}_if_i^{(0)}d{\bf{c}}_i,\quad
n_iT_i=\frac{1}{3}m_i\int{\bf{C}}_i^2f_i^{(0)}d{\bf{c}}_i.\\
\end{split}
\label{a18}
\end{equation}
The constitutive relations such as the solid stress tensor ${\bf{p}}_i$, the solid-solid drag force between different species ${\bm{\mathcal{F}}}_{ij}$, the heat flux ${\bf{q}}_i$, and the energy dissipation rate $\mathcal{N}_{d,i}$ of species $i$ in the hydrodynamic equations can be expanded as the following forms:
\begin{equation}
\begin{split}
{\bf{p}}_i={\bf{p}}_i^{O(1)}+{\bf{p}}_i^{O(K)}+\cdots\\
\end{split}
\label{a19}
\end{equation}
\begin{equation}
\begin{split}
{\bm{\mathcal{F}}}_{Dij}={\bm{\mathcal{F}}}_{Dij}^{O(1)}+{\bm{\mathcal{F}}}_{Dij}^{O(K)}+\cdots\\
\end{split}
\label{a20}
\end{equation}
\begin{equation}
\begin{split}
{\bf{q}}_i={\bf{q}}_i^{O(1)}+{\bf{q}}_i^{O(K)}+\cdots\\
\end{split}
\label{a21}
\end{equation}
\begin{equation}
\begin{split}
\mathcal{N}_{d,i}=\mathcal{N}_{d,i}^{O(1)}+\mathcal{N}_{d,i}^{O(K)}+\cdots\\
\end{split}
\label{a22}
\end{equation}
The subscript $O(1)$ signifies the constitutive relations related to the Maxwellian PVDF, while $O(K)$ denotes those related to the perturbation function. It is worth stressing that the above results derived by the single-parameter Chapman-Enskog method cannot differentiate between the factors that cause the system to derive from the equilibrium state. These factors may include variations in the hydrodynamic field, different species and inelastic interactions. However, all N-S order terms have been included, thus, from another perspective, this approach can well investigate the impact of cross-effect. Actually, it has some advantages in improving the prediction accuracy of the multifluid model for polydisperse gas-solid flow system \citep{zhao2021kinetic,shi2022critical}.

\subsection{Navier-Stokes order constitutive relations}
Referring to the Eq.(\ref{a6})-Eq.(\ref{a9}), the integrants are not only related to the PVDF
but also to the relationship between the pre- and post-velocities or energy. For the complete and analytical calculation of complex integrals, the variable replacement proposed by the previous works has been used in expressing the relation between pre- and post-velocities or energy as follows \citep{zhao2020note,zhao2021kinetic}:
\begin{equation}
\begin{split}
{\bf{c}}_i^{'}-{\bf{c}}_i=\eta_i({\bf{g}}\cdot{\bf{k}}){\bf{k}}
\end{split}
\label{a23}
\end{equation}
\begin{equation}
\begin{split}
{\bf{c}}_i^{'2}-{\bf{c}}_i^2
=\bigg[\eta_i^2-2\eta_i\frac{\theta_i}{\theta_i+\theta_j}\bigg]({\bf{k}}\cdot{\bf{g}})^2+2\eta_i({\bf{k}}\cdot{\bf{g}})({\bf{k}}\cdot{\bf{G}}^{\#}),
\end{split}
\label{a24}
\end{equation}
where the relevant variable replacement (${\bf{g}}$, ${\bf{G}}^{\#}$) is
\begin{equation}
\left\{
\begin{aligned}
&{\bf{g}}={\bf{c}}_j-{\bf{c}}_i\\
&{\bf{G}}^{\#}=\frac{m_iT_j{\bf{c}}_i+m_jT_i{\bf{c}}_j}{m_iT_j+m_jT_i},
\end{aligned}
\right.
\label{a25}
\end{equation}
where $\theta_i\equiv\frac{T_i}{m_i}$ and $\eta_i\equiv\frac{m_j}{m_0}(1+e_{ij})$,  {{$e_{ij}\equiv\frac{1}{2}(e_i+e_j)$}} is the restitutive coefficient of inelastic collision between particle $i$ and particle $j$.

Substituting these relations between pre- and post-collisions (\ref{a23}) and (\ref{a24}), and the PVDF (\ref{a10}) into the Eq.(\ref{a6})-Eq.(\ref{a9}), the solid stress tensor, the solid-solid drag force, the heat flux and the energy dissipation rate can be deduced one by one as reported in the following parts. {{The technical details are similar to our previous work \citep{zhao2021kinetic}, for the sake of brevity, we only provided the final results and not included derivation details of all constitutive relations. }}
\subsubsection{Solid stress tensor}
The solid stress tensor can be expressed as two parts:
\begin{equation}
\begin{split}
{\bf{p}}_i={\bf{p}}_i^{O(1)}+{\bf{p}}_i^{O(K)}.\\
\end{split}
\label{a26}
\end{equation}
The first term of solid stress tensor ${\bf{p}}_i^{O(1)}$ is
\begin{equation}
\begin{split}
&{\bf{p}}_i^{O(1)}\\
=&m_i\int{\bf{C}}_i{\bf{C}}_if_i^{(0)}d{\bf{c}}\\
+&\sum_j^N\bigg\{\frac{d_{ij}^3}{2}g_{ij}m_i\int\int\int\limits_{({\bf{g}}\cdot{\bf{k}})>0} ({\bf{c}}_i^{'}-{\bf{c}}_i)f_i^{(0)}f_j^{(0)}{\bf{k}}({\bf{k}}\cdot{\bf{g}})d{\bf{k}}d{\bf{c}}_id{\bf{c}}_j\\
+&\frac{d_{ij}^4}{4}g_{ij}m_i\int\int\int\limits_{({\bf{g}}\cdot{\bf{k}})>0}
({\bf{c}}_i^{'}-{\bf{c}}_i)f_i^{(0)}f_j^{(0)}\nabla\ln\frac{f_i^{(0)}}{f_j^{(0)}}\cdot({\bf{k}}\cdot{\bf{g}}){\bf{kk}}d{\bf{k}}d{\bf{c}}_id{\bf{c}}_j\bigg\}\\
=&m_in_i\theta_i{\bf{I}}\\
&+\sum_j^N\bigg\{\\
&\frac{\pi}{15}(1+e_{ij})d_{ij}^3g_{ij}\frac{m_im_jn_in_j}{m_0}
\bigg[
[5(\theta_i+\theta_j)+({\bf{u}}_j-{\bf{u}}_i)^2]{\bf{I}}+
2({\bf{u}}_j-{\bf{u}}_i)({\bf{u}}_j-{\bf{u}}_i)\bigg]\\
+&\frac{1}{3}\times 2^{-3}(1+e_{ij}){\pi}^{-\frac{1}{2}}d_{ij}^4g_{ij}(\frac{m_im_j}{m_0}) n_in_j(\theta_i+\theta_j)\exp(-|\bar{{\bf{g}}^{*}}|^2)\\
\times&\bigg[
\bigg(h_1(\overline{g^{*}})\frac{1}{({\overline{{{g}}^{*}}})^3}({\bf{w}}_1\cdot\overline{{\bf{g}}^*})(\overline{{\bf{g}}^*}\,\overline{{\bf{g}}^*})
+h_2(\overline{g^*})\frac{1}{\overline{g^*}}\bigg(2\mathring{\overline{\overline{{\bf{w}}_1\overline{{\bf{g}}^*}}}}+
\frac{5}{3}({\bf{w}}_1\cdot\overline{{\bf{g}}^*}){\bf{I}}\bigg)
+h_3(\overline{g^{*}})({\bf{w}}_1\cdot{\overline{{\bf{g}}^*}}){\bf{I}}
+h_3(\overline{g^{*}})({\bf{w}}_1\overline{{\bf{g}}^{*}}+\overline{{\bf{g}}^{*}}{\bf{w}}_1)]\bigg)\\
+&2(\theta_{i}+\theta_{j})\bigg(h_4(\overline{g^{*}})({\bf{w}}_3\cdot\overline{{\bf{g}}^*})(\overline{{\bf{g}}^*}\,\overline{{\bf{g}}^*})
+h_5(\overline{g^*})\bigg(2\mathring{\overline{\overline{{\bf{w}}_3\overline{{\bf{g}}^*}}}}+
\frac{5}{3}({\bf{w}}_3\cdot\overline{{\bf{g}}^*}){\bf{I}}\bigg)
+H_6(\overline{g^{*}})({\bf{w}}_3\cdot{\overline{{\bf{g}}^*}}){\bf{I}}
+H_6(\overline{g^{*}})({\bf{w}}_3\overline{{\bf{g}}^{*}}+\overline{{\bf{g}}^{*}}{\bf{w}}_3)]\bigg)\\
+&[2(\theta_{i}+\theta_{j})]^{\frac{1}{2}}
\bigg(H_7(\overline{g^*})({\bf{W}}_4:\overline{{\bf{g}}^*}\,\overline{{\bf{g}}^*})\overline{{\bf{g}}^*}\,\overline{{\bf{g}}^*}
+H_8(\overline{g^*})[\frac{1}{4}\bigg(tr({\bf{W}}_4){\bf{I}}+({\bf{W}}_4+{\bf{W}}_4^{T})\bigg)]
+H_9(\overline{g^*})K(W_{4ij}{\overline{g_{lk}^*}})\\
+&\pi[M_1(\overline{g^{*}})tr({\bf{W}}_4){\bf{I}}+M_2(\overline{g^*})({\bf{W}}_4:\overline{{\bf{g}}^*}\,\overline{{\bf{g}}^*}){\bf{I}}]
+\pi[h_6(\overline{g^{*}})({\bf{W}}_4+{\bf{W}}_4^T)
+h_7(\overline{g^{*}})[({\bf{W}}_4\cdot\overline{{\bf{g}}^{*}}){\overline{{\bf{g}}^{*}}}
+{\overline{{\bf{g}}^{*}}}({\bf{W}}_4\cdot\overline{{\bf{g}}^{*}})]\bigg)\bigg]
\\
&\bigg\},
\label{a27}
\end{split}
\end{equation}
where some relevant coefficient functions, vectors and second order tensor are listed as:
\begin{equation}
\begin{split}
h_1&=\frac{\overline{g^*}}{2}H_9=
\frac{\pi}{8{\overline{g^*}}^4}[30\overline{g^*}-16\overline{g^*}^3+8\overline{g^*}^5+\sqrt{\pi}(-15+18\overline{g^*}^2-12\overline{g^*}^4+8\overline{g^*}^6)erf(\overline{g^*})\exp(\overline{g^*}^2)]\\
h_2&=\frac{\overline{g^*}}{2}H_8=
\frac{\pi}{8{\overline{g^*}}^4}[-6\overline{g^*}+4\overline{g^*}^3+\sqrt{\pi}(3-4\overline{g^*}^2+4\overline{g^*}^4)erf(\overline{g^*})\exp(\overline{g^*}^2)]\\
M_1&=h_6=\frac{1}{2\pi}h_3=
\frac{1}{8\overline{g^*}^3}[2(\overline{g^*}+2{\overline{g^{*}}}^3)
+\sqrt{\pi}(-1+4{\overline{g^{*}}}^2+4{\overline{g^{*}}}^4)erf(\overline{g^{*}})\exp({\overline{g^{*}}}^2)]\\
M_2&=h_7=2{\overline{g^{*}}}^2h_5
=\frac{1}{8{\overline{g^{*}}}^3}[-6\overline{g^{*}}+8\overline{g^{*}}^3+8\overline{g^{*}}^5+
\sqrt{\pi}(3-6\overline{g^{*}}^2+12\overline{g^{*}}^4+8\overline{g^{*}}^6)erf(\overline{g^{*}})\exp(\overline{g^{*}}^2)]\\
h_4&=\frac{\pi}{16\overline{g^{*}}^7}[2\overline{g^{*}}(15-4\overline{g^{*}}^2+12\overline{g^{*}}^4+8\overline{g^{*}}^6)
+\sqrt{\pi}(-15+24\overline{g^{*}}^2-24\overline{g^{*}}^4+32\overline{g^{*}}^6+16\overline{g^{*}}^8)erf(\overline{g^{*}})\exp({\overline{g^{*}}}^2)]\\
H_6&=\frac{\pi}{8\overline{g^{*}}^3}[6\overline{g^{*}}+32\overline{g^{*}}^3+8\overline{g^{*}}^5+\sqrt{\pi}(-3+18\overline{g^{*}}^2+36\overline{g^{*}}^4+8\overline{g^{*}}^6)erf(\overline{g^{*}})\exp(\overline{g^{*}}^2)]\\
H_7&=\frac{\pi}{4\overline{g^{*}}^5}
[2\overline{g^{*}}(-105+50\overline{g^{*}}^2-20\overline{g^{*}}^4+8\overline{g^{*}}^6)
+\sqrt{\pi}(105+8\overline{g^{*}}^2(-15+9\overline{g^{*}}^2-4\overline{g^{*}}^4+2\overline{g^{*}}^6))erf(\overline{g^{*}})\exp(\overline{g^{*}}^2)]
\\
{\bf{w}}_1&=\nabla [\ln n_{i}-\ln n_{j}+\frac{3}{2}(\ln\theta_{j}-\ln \theta_{i})]
+\frac{{\bf{u}}_{j}-{\bf{u}}_{i}}{\theta_{j}+\theta_{i}}\cdot (\nabla {\bf{u}}_{j}+\nabla{\bf{u}}_{i})
-\frac{3}{2(\theta_{i}+\theta_{j})\theta_{i}\theta_{j}}(\theta_{i}^2\nabla\theta_{j}-\theta_{j}^2\nabla\theta_{i})
-\frac{({\bf{u}}_{j}-{\bf{u}}_{i})^2}{2(\theta_{j}+\theta_{i})^2}(\nabla\theta_{j}-\nabla\theta_{i})\\
{\bf{w}}_3&=\frac{\nabla\theta_{i}-\nabla\theta_{j}}{2(\theta_{j}+\theta_{i})^2}\\
{\bf{W}}_4&={-\frac{\nabla{\bf{u}}_{j}+\nabla{\bf{u}}_{i}}{\theta_{i}+\theta_{j}}
}-\bigg[\frac{\nabla\theta_{j}-\nabla\theta_{i}}{(\theta_{i}+\theta_{j})^2}\bigg]({\bf{u}}_{i}-{\bf{u}}_{j})\,.
\label{a28}
\end{split}
\end{equation}
where the dimensionless relative velocity between species $i$ and species $j$ is denoted as $\bar{\bf{g}}^{*}\equiv[\frac{1}{2(\theta_i+\theta_j)}]^{\frac{1}{2}}({\bf{u}}_j-{\bf{u}}_i)$, and its modulus is denoted as $\bar{g}^{*}\equiv|\bar{\bf{g}}^{*}|=[\frac{1}{2(\theta_i+\theta_j)}]^{\frac{1}{2}}|{\bf{u}}_j-{\bf{u}}_i|$. Moreover, the error function is defined as $erf(\bar{g}^{*})=\frac{2}{\sqrt{\pi}}\int_0^{\bar{g}^{*}}\exp(-y^2)dy$. {{The $K(W_{4ij}g_{sk}^*)$ is a symmetric matrix and its components are written as (where the line number and column number are denoted as $\mathcal{I}$ and $\mathcal{J}$, respectively)
\begin{equation}
\begin{split}
&\mathcal{I}=1,\, \mathcal{J}=1:\notag\\
&(\frac{3}{2}W_{4xx}+\frac{1}{4}W_{4yy}+\frac{1}{4}W_{4zz}){\overline{g_{x}^*}}^2
+\frac{1}{4}W_{4yy}{\overline{g_{y}}}^2+\frac{1}{4}W_{4zz}{\overline{g_{z}^*}}^2
\notag\\
&+\frac{3}{4}(W_{4xy}+W_{4yx}){\overline{g_{x}^*}}\,{\overline{g_{y}^*}}
+\frac{3}{4}(W_{4xz}+W_{4zx}){\overline{g_{x}^*}}\,{\overline{g_{z}^*}}
+\frac{1}{4}(W_{4yz}+W_{4zy}){\overline{g_{z}^*}}\,{\overline{g_{y}^*}}\notag\\
\notag\\
&\mathcal{I}=1,\, \mathcal{J}=2\,\&\,\mathcal{I}=2,\, \mathcal{J}=1:\notag\\
&\frac{1}{4}(W_{4xy}+W_{4yx}){\overline{g_{x}^*}}^2
+\frac{1}{4}(W_{4xy}+W_{4yx}){\overline{g_{y}^*}}^2\notag\\
&
+(\frac{3}{4}W_{4xx}+\frac{3}{4}W_{4yy}+\frac{1}{4}W_{4zz}){\overline{g_{x}^*}}\,{\overline{g_{y}^*}}
+\frac{1}{4}(W_{4yz}+W_{4zy}){\overline{g_{x}^*}}\,{\overline{g_{z}^*}}
+\frac{1}{4}(W_{4xz}+W_{4zx}){\overline{g_{y}^*}}\,{\overline{g_{z}^*}}\notag\\
\notag\\
&\mathcal{I}=1,\, \mathcal{J}=3\,\&\,\mathcal{I}=3,\, \mathcal{J}=1:\notag\\
&\frac{1}{4}(W_{4xz}+W_{4zx}){\overline{g_{x}^*}}^2
+\frac{1}{4}(W_{4xz}+W_{4zx}){\overline{g_{z}^*}}^2\notag\\
&
+\frac{1}{4}(W_{4yz}+W_{4zy}){\overline{g_{x}^*}}\,{\overline{g_{y}^*}}
+(\frac{3}{4}W_{4xx}+\frac{3}{4}W_{4zz}+\frac{1}{4}W_{4yy}){\overline{g_{x}^*}}\,{\overline{g_{z}^*}}
+\frac{1}{4}(W_{4xy}+W_{4yx}){\overline{g_{y}^*}}\,{\overline{g_{z}^*}}\notag\\
\notag\\
&\mathcal{I}=2,\, \mathcal{J}=2:\notag\\
&\frac{1}{4}W_{4xx}{\overline{g_{x}^*}}^2
+(\frac{3}{2}W_{4yy}+\frac{1}{4}W_{4xx}+\frac{1}{4}W_{4zz}){\overline{g_{y}^*}}^2
+\frac{1}{4}W_{4zz}{\overline{g_{z}}}^2\notag\\
&\frac{3}{4}(W_{4yx}+W_{4xy}){\overline{g_{x}^*}}\,{\overline{g_{y}^*}}
+\frac{1}{4}(W_{4xz}+W_{4zx}){\overline{g_{x}^*}}\,{\overline{g_{z}^*}}
+\frac{3}{4}(W_{4yz}+W_{4zy}){\overline{g_{z}^*}}\,{\overline{g_{y}^*}}\notag\\
\notag\\
&\mathcal{I}=2,\, \mathcal{J}=3\,\&\,\mathcal{I}=3,\, \mathcal{J}=2:\notag\\
&\frac{1}{4}(W_{4yz}+W_{4zy}){\overline{g_{y}^*}}^2
+\frac{1}{4}(W_{4yz}+W_{4zy}){\overline{g_{z}^*}}^2\notag\\
&+\frac{1}{4}(W_{4xz}+W_{4zx}){\overline{g_{x}^*}}\,{\overline{g_{y}^*}}
+\frac{1}{4}(W_{4xy}+W_{4yx}){\overline{g_{x}^*}}\,{\overline{g_{z}^*}}
+(\frac{3}{4}W_{4yy}+\frac{3}{4}W_{4zz}+\frac{1}{4}W_{4xx}){\overline{g_{y}^*}}\,{\overline{g_{z}^*}}\notag\\
\notag\\
&\mathcal{I}=3,\, \mathcal{J}=3:\notag\\
&\frac{1}{4}W_{4xx}{\overline{g_{x}^*}}^2
+\frac{1}{4}W_{4yy}{\overline{g_{y}^*}}^2
+(\frac{1}{4}W_{4xx}+\frac{1}{4}W_{4yy}+\frac{3}{2}W_{4zz}){\overline{g_{z}^*}}^2
\notag\\
&+\frac{1}{4}(W_{4xy}+W_{4yx}){\overline{g_{x}^*}}\,{\overline{g_{y}^*}}
+\frac{3}{4}(W_{4xz}+W_{4zx}){\overline{g_{x}^*}}\,{\overline{g_{z}^*}}
+\frac{3}{4}(W_{4yz}+W_{4zy}){\overline{g_{y}^*}}\,{\overline{g_{z}^*}}\,.\\
\label{a28_1}
\end{split}
\end{equation}
}}
The second part of solid stress tensor is
\begin{equation}
\begin{split}
&{\bf{p}}_i^{O(K)}\\
=&m_i\int{\bf{C}}_i{\bf{C}}_if_i^{(0)}\Phi_i^{(K)}d{\bf{c}}\\
+&\sum_j^N\bigg\{\frac{d_{ij}^3}{2}g_{ij}\int \int\int\limits_{({\bf{g}}\cdot{\bf{k}})>0} m_i({\bf{c}}_i^{'}-{\bf{c}})f_i^{(0)}f_j^{(0)}(\Phi_i^{(K)}+\Phi_j^{(K)}){\bf{k}}({\bf{k}}\cdot{\bf{g}})d{\bf{k}}d{\bf{c}}_id{\bf{c}}_j\bigg\}\\
=&2[-A_iD_i]m_in_i\theta_i^2\mathring{\overline{\overline{\nabla{\bf{u}}_i}}}\\
+&\sum_j^N\bigg\{
-\frac{4\pi}{15}(1+e_{ij})d_{ij}^3g_{ij}\frac{m_im_j}{m_0}n_in_j[A_iD_i\theta_i^2\mathring{\overline{\overline{\nabla{\bf{u}}_i}}}
+A_jD_j\theta_j^2\mathring{\overline{\overline{\nabla{\bf{u}}_j}}}]\bigg\}.
\label{a29}
\end{split}
\end{equation}

Referring to the final form of the solid stress tensor, we have found that: (i) the pressure of species $i$ not only relates to itself granular temperature, but also relates to the granular temperature of particles interacting  with it. Meanwhile, the square of the slip velocity between different species as the fluctuation energy at the higher level has been included into; (ii) more driving forces make contribution to the shear stress tensor besides velocity gradient, such as the differences in volume fraction gradients and granular temperature gradients between different species. The results indicate that in multiphase flow modeling of polydisperse systems, the interaction of particles from different species cannot be ignored.

\subsubsection{Solid-solid drag force}
The drag force between solid particles, which appears as the momentum source term in the momentum conservation equation, can be divided into two parts:
\begin{equation}
\begin{split}
{\bm{\mathcal{F}}}_{Dij}={\bm{\mathcal{F}}}_{Dij}^{O(1)}+{\bm{\mathcal{F}}}_{Dij}^{O(K)},\\
\end{split}
\label{a30}
\end{equation}
where the first part is
\begin{align}
&{\bm{\mathcal{F}}}_{D,ij}^{O(1)}\notag\\
=&d_{ij}^2g_{ij}m_i\int\int\int\limits_{({\bf{g}}\cdot{\bf{k}})>0}
({\bf{c}}_i^{'}-{\bf{c}}_i)f_i^{(0)}f_j^{(0)}({\bf{g}}\cdot{\bf{k}})d{\bf{k}}d{\bf{c}}_id{\bf{c}}_j\notag\\
&+\frac{d_{ij}^3}{2}g_{ij}m_i\int\int\int\limits_{({\bf{g}}\cdot{\bf{k}})>0}
({\bf{c}}_i^{'}-{\bf{c}}_i)f_i^{(0)}f_j^{(0)}\nabla\ln\frac{f_i^{(0)}}{f_j^{(0)}}\cdot{\bf{k}}({\bf{g}}\cdot{\bf{k}})
d{\bf{k}}d{\bf{c}}_id{\bf{c}}_j\notag\\
=&{(1+e_{ij})d_{ij}^2}g_{ij}\frac{m_im_j}{m_0}n_in_j(\theta_i+\theta_j)\pi^{\frac{1}{2}}
\bigg[(1+\frac{1}{2\bar{{g}}^{*2}})\exp(-\bar{{g}}^{*2})
+\sqrt{\pi}(\frac{1}{\bar{g}^{*}}+\bar{g}^{*}
-\frac{1}{4\bar{g}^{*3}})erf(\bar{g}^{*})\bigg]\bar{\bf{g}}^{*}\notag\\
+&\frac{2(1+e_{ij})\pi}{15}d_{ij}^3g_{ij}\frac{m_im_j}{m_0} n_in_j(\theta_i+\theta_j)\notag\\
\times&\bigg\{\frac{1}{2}(3+2{\overline{g^*}}^2){\bf{w}}_1
+2^{\frac{1}{2}}(\theta_{i}+\theta_{j})^{\frac{1}{2}}[\frac{1}{2}(5+2{\overline{g^*}}^2)]{\bf{W}}_4\cdot{\overline{{\bf{g}}^*}}
+2(\theta_{i}+\theta_{j})[\frac{1}{4}(15+4{\overline{g^*}}^2(5+{\overline{g^*}}^2))]{\bf{w}}_3\notag\\
+&2\bigg[[\frac{1}{2}{\bf{I}}+\frac{1}{2(\theta_{i}+\theta_{j})}\overline{{\bf{g}}}\,\overline{{\bf{g}}}]\cdot{\bf{w}}_1
+[({\bf{W}}_4\cdot{\overline{{\bf{g}}}}+{\overline{{\bf{g}}}}\cdot{\bf{W}}_4)
+{\overline{{\bf{g}}}}\cdot tr({\bf{W}}_4){\bf{I}}
+\frac{1}{2(\theta_{i}+\theta_{j})}{\overline{{\bf{g}}}}\,{\overline{{\bf{g}}}}\cdot({\bf{W}}_4\cdot{\overline{{\bf{g}}}})]\notag\\
+&2(\theta_{i}+\theta_{j})[\frac{1}{4}(5+2{\overline{g^*}}^2){\bf{w}}_3
+\frac{1}{2}(7+2{\overline{g^*}}^2){\overline{{\bf{g}}^*}}\,{\overline{{\bf{g}}^*}}\cdot{\bf{w}}_3\bigg]\bigg\},
\label{a31}
\end{align}
where the notations are $\mathring{h}_5=\pi h_5$, $\bar{\bf{G}}^{\#}\equiv\frac{\theta_i{\bf{u}}_j+\theta_j{\bf{u}}_i}{\theta_i+\theta_j}$,  ${\bf{R}}_i\equiv\bar{\bf{G}}^{\#}-{\bf{u}}_i=\frac{\theta_i({\bf{u}}_j-{\bf{u}}_i)}{\theta_i+\theta_j}$,
 and ${\bf{R}}_j\equiv\bar{\bf{G}}^{\#}-{\bf{u}}_j=\frac{\theta_j({\bf{u}}_i-{\bf{u}}_j)}{\theta_i+\theta_j}$.

The first part's solid-solid drag force form is remarkably complex due to the analytical solving of the addition of $f_i^{(0)}f_j^{(0)}\nabla\ln \frac{f_i^{(0)}}{f_j^{(0)}}$ contribution in conjunction with $f_i^{(0)}f_j^{(0)}$. Additionally, once we combine the definitions of ${\bf{w}}_1$, ${\bf{w}}_3$, and ${\bf{W}}_3$ from Eq.(\ref{a28}), we can summarize the cause of the complex drag force as: in the polydisperse system, the driving force of segregation is not only the slip velocity of different species $({\bf{u}}_i-{\bf{u}}_j)$, but also  the gradient differences of macroscopic hydrodynamic variables such as $(\nabla n_i-\nabla n_j)$, $(\nabla {\bf{u}}_i-\nabla {\bf{u}}_j)$ and $(\nabla \theta_i-\nabla\theta_j)$. Moreover, the solid-solid drag force coefficients  vary based on the driving force.
%In terms of these two sides, the complex first part of the solid-solid drag force is obtained as Eq.(\ref{a31}). In other words, the comprehensive and precise solid-solid drag force has been expanded to include additional driving forces and drag coefficients. By referring to the work of \cite{duan_yu_chen_zhou_2022}, they have found the part of ${\bm{\mathcal{F}}}_{D,ij}^{O(1)}$ only associated with the $f_i^{(0)}$ is modified by the effective relative velocity, and the multifluid model coupled with it has a strong predictive capability. Appropriate corrections for the driving force and the drag coefficient will improve the prediction accuracy of the multifluid model.

The second part of solid-solid drag force is
\begin{align}
&{\bm{\mathcal{F}}}_{D,ij}^{O(K)}\notag\\
=&\sum_j^N\bigg\{d_{ij}^2g_{ij}m_i\int\int\int\limits_{({\bf{g}}\cdot{\bf{k}})>0}
({\bf{c}}_i^{'}-{\bf{c}}_i)f_i^{(0)}f_j^{(0)}(\Phi_i^{(K)}+\Phi_j^{(K)})({\bf{g}}\cdot{\bf{k}})d{\bf{k}}d{\bf{c}}_id{\bf{c}}_j\bigg\}\notag\\
=&\sum_j^N\bigg\{\notag\\
&(1+e_{ij})d_{ij}^2g_{ij}\frac{m_im_j}{m_0}n_in_j
\pi^{-\frac{1}{2}}\exp(-\bar{g^*}^2)\notag\\
\times&\bigg\{[-A_iB_i]\bigg[2\theta_i\theta_j\bigg(\frac{5}{4\theta_i}h_3
+\frac{1}{2\theta_j}H_6+(\frac{\theta_i+\theta_j}{2\theta_i\theta_j})h_3(\frac{{\bf{R}}_i^2}{2\theta_i}-\frac{5}{2})\bigg)\bar{{\bf{g}}^{*}}{\bf{R}}_i\notag\\
+&2^{\frac{3}{2}}(\theta_i+\theta_j)^{\frac{1}{2}}
(\theta_i\theta_j)\pi(h_6{\bf{I}}+h_7\frac{\bar{{\bf{g}}^{*}}{\bar{\bf{g}}^{*}}}{\bar{g^{*}}^2})\bigg[
\bigg(\frac{5}{4\theta_j}-\frac{1}{4\theta_i\theta_j}{\bf{R}}_i^2-\frac{5}{4(\theta_i+\theta_j)}\bigg){\bf{I}}
-\frac{1}{2\theta_i\theta_j}{\bf{R}}_i{\bf{R}}_i\bigg]\notag\\
-&2^{-\frac{1}{2}}(\theta_i+\theta_j)^{-\frac{1}{2}}
\theta_i^2[H_6{\bf{I}}+{\mathcal{H}}_8\bar{{\bf{g}}}^{*}\bar{{\bf{g}}}^{*}]\notag\\
+&2\theta_i[h_4\bar{{\bf{g}}}^{*}(\bar{{\bf{g}}}^{*}\bar{{\bf{g}}}^{*}\cdot{\bf{R}}_i)
+{\mathring{h}}_5(\bar{{\bf{g}}}^{*}{\bf{R}}_i+{\bf{R}}_i\bar{{\bf{g}}}^{*}+(\bar{{\bf{g}}}^{*}\cdot{\bf{R}}_i){\bf{I}})]\bigg]
\cdot\nabla\ln\theta_i\notag\\
+&[-A_jB_j]\bigg[2\theta_i\theta_j\bigg(\frac{5}{4\theta_j}h_3
+\frac{1}{2\theta_i}H_6+(\frac{\theta_i+\theta_j}{2\theta_i\theta_j})h_3(\frac{{\bf{R}}_j^2}{2\theta_j}-\frac{5}{2})\bigg)\bar{{\bf{g}}^{*}}{\bf{R}}_j\notag\\
+&2^{\frac{3}{2}}(\theta_i+\theta_j)^{\frac{1}{2}}
(\theta_i\theta_j)\pi(h_6{\bf{I}}+h_7\frac{\bar{{\bf{g}}^{*}}{\bar{\bf{g}}^{*}}}{\bar{g^{*}}^2})\bigg[
\bigg(-\frac{5}{4\theta_i}+\frac{1}{4\theta_i\theta_j}{\bf{R}}_j^2+\frac{5}{4(\theta_i+\theta_j)}\bigg){\bf{I}}
+\frac{1}{2\theta_i\theta_j}{\bf{R}}_j{\bf{R}}_j\bigg]\notag\\
+&2^{-\frac{1}{2}}(\theta_i+\theta_j)^{-\frac{1}{2}}
\theta_j^2[H_6{\bf{I}}+{\mathcal{H}}_8\bar{{\bf{g}}}^{*}\bar{{\bf{g}}}^{*}]\notag\\
+&2\theta_j[h_4\bar{{\bf{g}}}^{*}(\bar{{\bf{g}}}^{*}\bar{{\bf{g}}}^{*}\cdot{\bf{R}}_j)
+{\mathring{h}}_5(\bar{{\bf{g}}}^{*}{\bf{R}}_j+{\bf{R}}_j\bar{{\bf{g}}}^{*}+(\bar{{\bf{g}}}^{*}\cdot{\bf{R}}_j){\bf{I}})]\bigg]
\cdot\nabla\ln\theta_j\bigg\}\notag\\
+&2(1+e_{ij})d_{ij}^2g_{ij}\frac{m_im_j}{m_0}n_in_j\pi^{-\frac{1}{2}}\exp(-{\bar{g}}^{*2})(\theta_i+\theta_j)^2\notag\\
\times&\bigg\{
[-A_iD_i]\bigg[\bigg(\frac{1}{2(\theta_i+\theta_j)}h_3\bar{{\bf{g}}}^{*}({\bf{R}}_i{\bf{R}}_i-\frac{1}{3}{\bf{R}}_i^2{\bf{I}})
-H_6\bar{{\bf{g}}}^{*}(\frac{\theta_i^2}{3(\theta_i+\theta)j)^2}){\bf{I}}\bigg):\nabla{\bf{u}}_i\notag\\
-&2^{-\frac{1}{2}}\theta_i(\theta_i+\theta_j)^{-\frac{3}{2}}
\bigg(h_6[{\bf{R}}_i\cdot(\nabla{\bf{u}}_i+\nabla{\bf{u}}_i^{T})
-\frac{2}{3}{\bf{R}}_i(\nabla\cdot{\bf{u}}_i)]
+\frac{h_7}{{\bar{g}}^{*2}}[{\bar{\bf{g}}}^{*}({\bf{R}}_i{\bar{\bf{g}}}^{*}
+{\bar{\bf{g}}}^{*}{\bf{R}}_i)-\frac{2}{3}({\bf{R}}_i\cdot{{\bar{\bf{g}}}^{*}{\bar{\bf{g}}}^{*}}){\bf{I}}]:\nabla{\bf{u}}_i\bigg)\notag\\
+&\theta_i^2(\theta_i+\theta_j)^{-2}
[h_4{\bar{\bf{g}}}^{*}({\bar{\bf{g}}}^{*}{\bar{\bf{g}}}^{*}:\nabla{\bf{u}}_i)
+{\mathring{h}}_5{\bar{\bf{g}}}^{*}\cdot\bigg((\nabla{\bf{u}}_i+{\nabla}{\bf{u}}_i^{T})+({\nabla}\cdot{\bf{u}}_i){\bf{I}}\bigg)]\bigg]\notag\\
+&[-A_jD_j]\bigg[\bigg(\frac{1}{2(\theta_i+\theta_j)}h_3{\bar{\bf{g}}}^{*}({\bf{R}}_j{\bf{R}}_j-\frac{1}{3}{\bf{R}}_j^2{\bf{I}})
-H_6{\bar{\bf{g}}}^{*}(\frac{\theta_j^2}{3(\theta_i+\theta)j)^2}){\bf{I}}\bigg):\nabla{\bf{u}}_j\notag\\
+&2^{-\frac{1}{2}}\theta_j(\theta_i+\theta_j)^{-\frac{3}{2}}
\bigg(h_6[{\bf{R}}_j\cdot(\nabla{\bf{u}}_j+\nabla{\bf{u}}_j^{T})
-\frac{2}{3}{\bf{R}}_j(\nabla\cdot{\bf{u}}_j)]
+\frac{h_7}{{\bar{g}}^{*2}}[{\bar{\bf{g}}}^{*}({\bf{R}}_j{\bar{\bf{g}}}^{*}
+{\bar{\bf{g}}}^{*}{\bf{R}}_j)-\frac{2}{3}({\bf{R}}_j\cdot{{\bar{\bf{g}}}^{*}{\bar{\bf{g}}}^{*}}){\bf{I}}]:\nabla{\bf{u}}_j\bigg)\notag\\
+&\theta_j^2(\theta_i+\theta_j)^{-2}
[h_4{\bar{\bf{g}}}^{*}({\bar{\bf{g}}}^{*}{\bar{\bf{g}}}^{*}:\nabla{\bf{u}}_j)
+{\mathring{h}}_5{\bar{\bf{g}}}^{*}\cdot\bigg((\nabla{\bf{u}}_j+{\nabla}{\bf{u}}_j^{T})+({\nabla}\cdot{\bf{u}}_j){\bf{I}}\bigg)]\bigg]\bigg\}\notag\\
&\bigg\},\notag\\
\label{a32}
\end{align}
where the coefficient function is
\begin{equation}
\begin{split}
\mathcal{H}_8&\equiv
\frac{\pi}{8{\bar{g^*}}^5}\bigg[2{\bar{g^*}}(-9+18{\bar{g^*}}^2+44{\bar{g^*}}^4+8{\bar{g^*}}^6)
+{{\exp({\bar{g^*}}^2)}}\sqrt{\pi}(9-24{\bar{g^*}}^2+72{\bar{g^*}}^4+96{\bar{g^*}}^6+16{\bar{g^*}}^8)erf({\bar{g^*}})\bigg].\\
\label{a33}
\end{split}
\end{equation}
The second part of the solid-solid drag force demonstrates the coupling effect of various species. This formula (\ref{a32}) yields a significant number of driving forces and drag coefficients.

\subsubsection{Heat flux}
By performing similar calculations for the solid stress tensor, it is possible to split the heat flux into two parts $
{\bf{q}}_i={\bf{q}}_i^{O(1)}+{\bf{q}}_i^{O(K)}
$, as represented by the flux term in the granular temperature equation:
\begin{align}
&{\bf{q}}_i^{O(1)}\notag\\
=&\frac{1}{2}m_i\int{\bf{C}}_i^2{\bf{C}}_if_i^{(0)}d{\bf{c}}\notag\\
+&\sum_j^N\bigg\{\frac{d_{ij}^3}{4}g_{ij}m_i\int\int\int\limits_{({\bf{g}}\cdot{\bf{k}})>0} ({\bf{C}}_i^{'2}-{\bf{C}}_i^2)f_i^{(0)}f_j^{(0)}{\bf{k}}({\bf{k}}\cdot{\bf{g}})d{\bf{k}}d{\bf{c}}_id{\bf{c}}_j\notag\\
+&\frac{d_{ij}^4}{8}g_{ij}m_i\int\int\int\limits_{({\bf{g}}\cdot{\bf{k}})>0}
({\bf{C}}_i^{'2}-{\bf{C}}_i^2)f_i^{(0)}f_j^{(0)}\nabla\ln\frac{f_i^{(0)}}{f_j^{(0)}}\cdot({\bf{k}}\cdot{\bf{g}}){\bf{kk}}d{\bf{k}}d{\bf{c}}_id{\bf{c}}_j\bigg\}\notag\\
=&0\notag\\
&+\sum_j^N\bigg\{\notag\\
&\frac{\pi}{20}d_{ij}^3m_ig_{ij}n_in_j
\bigg[[\eta_i^2-2\eta_i\frac{\theta_i}{\theta_i+\theta_j}][2(\theta_i+\theta_j)](5+2\bar{g}^{*2})\bar{\bf{g}}
+\frac{4\eta_i}{3}
(\bar{\bf{G}}^{\#}-{\bf{u}}_i)\cdot[2\bar{\bf{g}}\bar{\bf{g}}+(\frac{5}{2}B^{-1}{\bf{I}}+\bar{{g}}^{2}{\bf{I}})]\bigg]\notag\\
&+\frac{2^{-\frac{7}{2}}}{3}\pi^{\frac{1}{2}}d_{ij}^4m_ig_{ij}n_in_j[\eta_i^2-2\eta_i\frac{\theta_i}{\theta_i+\theta_j}]\notag\\
&\quad\times\bigg\{\bigg[(3\exp(-{\bar{g^*}}^2)h_6+\mathcal{L}_3){\bf{I}}
+3\exp(-{\bar{g^*}}^2)h_7\frac{{\overline{{\bf{g}}^{*}}}\,{\overline{{\bf{g}}^{*}}}}{{\bar{g^*}}^2}\bigg]
\cdot[{\bf{A}}_1+(\overline{{\bf{G}}^{\#}}\cdot{\bf{A}}_2)-(\frac{3}{2}A^{-1}+\overline{{G^{\#}}}^2){\bf{A}}_4]\notag\\
&\quad\quad-B^{-\frac{1}{2}}\exp(-{\bar{g^*}}^2)
\bigg[3h_4{\overline{{\bf{g}}^{*}}}\,{\overline{{\bf{g}}^{*}}}\cdot[(\overline{{\bf{G}}^{\#}}{\bf{A}}_6+{\bf{A}}_3)\cdot{{\overline{{\bf{g}}^{*}}}}]
+3h_4{\overline{{\bf{g}}^{*}}}\cdot[({\bf{A}}_3+\overline{{\bf{G}}^{\#}} {\bf{A}}_6)
+({\bf{A}}_3+\overline{{\bf{G}}^{\#}} {\bf{A}}_6)^{T}
+(({\bf{A}}_3+\overline{{\bf{G}}^{\#}} {\bf{A}}_6):{\bf{I}}){\bf{I}}]\notag\\
&\quad\quad\quad\quad\quad\quad\quad\quad\quad+\pi^{-1}H_6[{\overline{{\bf{g}}^{*}}}\cdot({\bf{A}}_3+\overline{{\bf{G}}^{\#}} {\bf{A}}_6)]\bigg]\notag\\
&\quad\quad-\pi^{-1}B^{-1}\exp(-{\bar{g^*}}^2)[(\frac{3}{2}H_6+{\mathcal{H}}_9){\bf{I}}+{\mathcal{H}}_8{\overline{{\bf{g}}^{*}}}\,{\overline{{\bf{g}}^{*}}}]
\cdot{\bf{A}}_5\bigg\}\notag\\
&+\frac{2^{-\frac{7}{2}}}{3}\pi^{\frac{1}{2}}d_{ij}^4m_ig_{ij}n_in_j[2\eta_i]
(\theta_i+\theta_j)^{\frac{3}{2}}\exp(-{\bar{g}}^{*2})\notag\\
&\quad\times\bigg\{\pi^{-1}
\bigg[
[{\overline{{\bf{g}}^{*}}}({\overline{{\bf{g}}^{*}}}{\overline{{\bf{g}}^{*}}}:{\bf{w}}_0)]
+{\overline{{\bf{g}}^{*}}}\cdot[{\bf{w}}_0+{\bf{w}}_0^{T}+({\bf{w}}_0:{\bf{I}}){\bf{I}}]\notag\\
&\quad\quad\quad\quad+B^{\frac{1}{2}}2\pi h_6[(\overline{{\bf{G}}^{\#}}\cdot{\overline{{\bf{g}}^{*}}})
\bigg({\bf{A}}_1-\frac{1}{2}A^{-1}(5+2A\overline{G^{\#}}^2){\bf{A}}_4\bigg)+
+{\overline{{\bf{g}}^{*}}}\bigg(\frac{1}{2}A^{-1}{\bf{A}}_2
+\overline{{\bf{G}}^{\#}} (\overline{{\bf{G}}^{\#}}\cdot{\bf{A}}_2)\bigg):{\bf{I}}]\notag\\
&\quad\quad\quad\quad-B^{-\frac{1}{2}}H_6(\overline{{\bf{G}}^{\#}}\cdot{\overline{{\bf{g}}^{*}}})\cdot{\bf{A}}_5\notag\\
&\quad\quad\quad\quad-[\frac{1}{4}H_7\frac{1}{{\bar{g^*}}^4}{\overline{{\bf{g}}^{*}}}\bigg(\overline{{\bf{G}}^{\#}} ({\overline{{\bf{g}}^{*}}}
\cdot{\bf{A}}_3):{\overline{{\bf{g}}^{*}}}\,{\overline{{\bf{g}}^{*}}}\bigg) +(\overline{{G^{\#}_{sx}}}+\overline{{G^{\#}_{sy}}}+\overline{{G^{\#}_{sz}}})H_9\frac{1}{{\bar{g^*}}^2}{\bm{\mathcal{H}_1}}
\notag\\
&\quad\quad\quad\quad+(\overline{{G^{\#}_{sx}}}+\overline{{G^{\#}_{sy}}}+\overline{{G^{\#}_{sz}}})\frac{1}{4}H_8
\bigg[(1,1,1)\cdot\bigg({\bf{A}}_3+2({\bf{A}}_3\cdot{\bf{I}})\bigg)\bigg]\notag\\
%%(3A_{311}+A_{321}+A_{331},A_{312}+3A_{322}+A_{332},A_{313}+A_{323}+3A_{333})\\
&\quad\quad\quad\quad+\pi h_6{\bf{A}}_3\cdot\overline{{\bf{G}}^{\#}} +\frac{\pi h_7}{{\bar{g^*}}^2}{\overline{{\bf{g}}^{*}}}(\overline{{\bf{G}}^{\#}} ({\overline{{\bf{g}}^{*}}}\cdot{\bf{A}}_3)):{\bf{I}}]\notag\\
&\quad\quad\quad\quad-A^{-1}\bigg(\frac{1}{2}{\bf{a}}_{61}+A{\bf{a}}_{62}
+\frac{1}{2}\pi({\bf{I}}+A\overline{{\bf{G}}^{\#}} \,\overline{{\bf{G}}^{\#}}):(h_6{\bf{I}}+\frac{h_7}{{\bar{g^*}}^2}{\overline{{\bf{g}}^{*}}}\,{\overline{{\bf{g}}^{*}}})
{\bf{A}}_6\bigg)\bigg]\notag\\
&\quad\quad\quad+\bigg[
(\overline{{\bf{G}}^{\#}}{\overline{{\bf{g}}^{*}}}+{\overline{{\bf{g}}^{*}}}\overline{{\bf{G}}^{\#}})
\cdot[2B^{\frac{1}{2}}h_6{\bf{A}}_1-A^{-1}B^{\frac{1}{2}}(5+2A\overline{{G^{\#}}}^2)h_6{\bf{A}}_4
-\pi^{-1}B^{-\frac{1}{2}}H_6{\bf{A}}_5]\notag\\
&\quad\quad\quad\quad{-}[\overline{{\bf{G}}^{\#}} (h_6{\bf{I}}+h_7\frac{{\overline{{\bf{g}}^{*}}}\,{\overline{{\bf{g}}^{*}}}}{{\bar{g^*}}^2}):{\bf{A}}_3
+h_6({\bf{A}}_3\cdot\overline{{\bf{G}}^{\#}})
+\frac{h_7}{{\bar{g}}^{*2}}{\overline{{\bf{g}}^{*}}}(\overline{{\bf{G}}^{\#}} {\overline{{\bf{g}}^{*}}}:{\bf{A}}_3)]\notag\\
&\quad\quad\quad\quad{-}[A^{-1}(h_6{\bf{I}}+\frac{h_7}{{\bar{g^*}}^2}{\overline{{\bf{g}}^{*}}}\,{\overline{{\bf{g}}^{*}}})
+\bigg(2h_6\overline{{\bf{G}}^{\#}} \,\overline{{\bf{G}}^{\#}} +\frac{h_7}{{\bar{g}}^{*2}}(\overline{{\bf{G}}^{\#}} \cdot{\overline{{\bf{g}}^{*}}})
(\overline{{\bf{G}}^{\#}}{\overline{{\bf{g}}^{*}}}+{\overline{{\bf{g}}^{*}}}\overline{{\bf{G}}^{\#}})\bigg)]\cdot{\bf{A}}_6\notag\\
&\quad\quad\quad\quad+A^{-1}B^{\frac{1}{2}}h_6[{\bf{A}}_2+({\bf{A}}_2:{\bf{I}})]\cdot{\overline{{\bf{g}}^{*}}}\bigg]\bigg\}
\notag\\
&\quad\quad\quad\quad-{\bf{u}}_{i}\cdot{\bf{p}}_{c2,ij}^{(K-c)}\notag\\
&\bigg\},
\label{a34}
\end{align}

\begin{equation}
\begin{split}
&{\bf{q}}_{i}^{O(K)}\\
=&\frac{1}{2}m_i\int{\bf{C}}_i^2{\bf{C}}_if_i^{(0)}\Phi_i^{(K)}d{\bf{C}}_i\\
+&\sum_j^N\bigg\{\frac{d_{ij}^3}{4}g_{ij}m_i\int\int\int\limits_{({\bf{g}}\cdot{\bf{k}})>0}
({\bf{C}}_i^{'2}-{\bf{C}}_i^2)f_i^{(0)}f_j^{(0)}(\Phi_i^{(K)}+\Phi_j^{(K)}){\bf{k}}({\bf{g}}\cdot{\bf{k}})d{\bf{k}}d{\bf{c}}_id{\bf{c}}_j\bigg\}\\
=&\frac{5}{2}m_in_i\theta_i^2[-A_iB_i]\cdot\nabla\ln\theta_i\\
+&\sum_j^N\bigg\{
d_{ij}^3m_ig_{ij}\pi n_in_j
\times\bigg[[-A_iB_i](\frac{1}{2}{\bf{I}}\cdot\nabla\ln\theta_i)
(2\eta_i-\eta_i^2)\theta_i^2
{+[-A_iD_i]{\frac{2}{5}
[\eta_i^2-\frac{4}{3}\eta_i](\bar{\bf{g}}
\cdot\mathring{\overline{\overline{\nabla{\bf{u}}_i}}})}\theta_i^2}\bigg]\\
&+d_{ij}^3m_ig_{ij}\pi n_in_j
\times\bigg[[-A_jB_j](\frac{1}{2}{\bf{I}}\cdot\nabla\ln\theta_j)
(\eta_i^2\theta_j^2)
+[-A_jD_j]\frac{1}{5}[2\eta_i^2\theta_j^2\bar{\bf{g}}\cdot
\mathring{\overline{\overline{\nabla{\bf{u}}_j}}}]
\bigg]\bigg\},\\
\label{a35}
\end{split}
\end{equation}
where ${\bf{p}}_{c2,ij}^{(K-c)}$ in Eq.(\ref{a34}) is the  part of ${\bf{p}}_{i}^{O(1)}$ related to the collision between particles from species $i$ and species $j$: ${\bf{p}}_{c2,ij}^{(K-c)}\equiv\frac{d_{ij}^4}{4}g_{ij}m_i\int\int\int\limits_{({\bf{g}}\cdot{\bf{k}})>0}
({\bf{c}}_i^{'}-{\bf{c}}_i)f_i^{(0)}f_j^{(0)}\nabla\ln\frac{f_i^{(0)}}{f_j^{(0)}}
\cdot({\bf{k}}\cdot{\bf{g}}){\bf{kk}}d{\bf{k}}d{\bf{c}}_id{\bf{c}}_j$
, its final form has been calculated analytical as Eq.(\ref{a27}).
{{Other coefficients related to above formula are
\begin{align}
A&=\frac{\theta_i+\theta_j}{2\theta_i\theta_j}\notag\\
B&=\frac{1}{2(\theta_i+\theta_j)}\notag\\
{\bf{A}}_1&
=\nabla[\ln n_{i}-\ln n_{j}+\frac{3}{2}(\ln \theta_{j}-\ln \theta_{i})]+
(\frac{{\bf{u}}_{j}}{\theta_{j}}\cdot\nabla{\bf{u}}_{j}{-}\frac{{\bf{u}}_{i}}{\theta_{i}}\cdot\nabla{\bf{u}}_{i})
+(\frac{u_{i}^2}{2\theta_{i}^2}\nabla\theta_{i}-\frac{u_{j}^2}{2\theta_{j}^2}\nabla\theta_{j})\notag\\
{\bf{A}}_2&
=\frac{1}{\theta_{i}}\nabla{\bf{u}}_{i}-\frac{1}{\theta_{j}}\nabla{\bf{u}}_{j}
+\frac{{\bf{u}}_{j}}{\theta_{j}^2}\nabla\theta_{j}{-}\frac{{\bf{u}}_{i}}{\theta_{i}^2}\nabla\theta_{i}\notag\\
{\bf{A}}_3&
=\frac{\nabla{\bf{u}}_{i}+\nabla{\bf{u}}_{j}}{\theta_{i}+\theta_{j}}
-\frac{{\bf{u}}_{j}}{\theta_{j}(\theta_{i}+\theta_{j})}\nabla\theta_{j}
-\frac{{\bf{u}}_{i}}{\theta_{i}(\theta_{i}+\theta_{j})}\nabla\theta_{i}\notag\\
{\bf{A}}_4&=\frac{1}{2\theta_{j}^2}\nabla\theta_{j}-\frac{1}{2\theta_{i}^2}\nabla\theta_{i}\notag\\
{\bf{A}}_5&=\frac{\nabla\theta_{j}-\nabla\theta_{i}}{2(\theta_{j}+\theta_{i})^2}\notag\\
{\bf{A}}_6&=\frac{\nabla\theta_{j}}{\theta_{j}(\theta_{j}+\theta_{i})}
+\frac{\nabla\theta_{i}}{\theta_{i}(\theta_{j}+\theta_{i})}\notag\\
{\bf{w}}_0&=B^{\frac{1}{2}}K_1{\bf{w}}-B^{-\frac{1}{2}}K_3\overline{{\bf{G}}^{\#}}{\bf{A}}_5
\notag\\
{{{\bf{w}}}}&={{\overline{{\bf{G}}^{\#}}{\bf{A}}_1+\frac{1}{2}A^{-1}{\bf{A}}_2
+\overline{{\bf{G}}^{\#}}(\overline{{\bf{G}}^{\#}}\cdot{\bf{A}}_2)
-\frac{1}{2}A^{-1}(5+2A\overline{{\bf{G}}^{\#}}^2)\overline{{\bf{G}}^{\#}}{\bf{A}}_4}}\notag\\
{{K_1}}&{{=\frac{1}{{\bar{g^*}}^3}h_1}}
\notag\\
{K_3}&={h_4} \notag\\
{\mathcal{H}}_9&\equiv
\frac{\pi}{8{\bar{g^*}}}[66{\bar{g^*}}+56{\bar{g^*}}^3+8{\bar{g^*}}^5
+{{\exp({\bar{g^*}}^2)}}\sqrt{\pi}(15+90{\bar{g^*}}^2+60{\bar{g^*}}^4+8{\bar{g^*}}^6)erf({\bar{g^*}})]\notag\\
{\bm{\mathcal{H}}}_1&\equiv(\mathcal{H}_{1x},\mathcal{H}_{1y},\mathcal{H}_{1z})
{=\bigg[({\overline{{\bf{g}}^{*}}}{\overline{{\bf{g}}^{*}}}\cdot{\bf{{A}}}_3)\cdot{\bf{I}}\bigg]\cdot(1,1,1)}\notag\\
\mathcal{H}_{1x}&
={\overline{{\bf{g}}^{*}}}_x^2(\frac{3}{2}A_{311}+\frac{1}{4}A_{321}+\frac{1}{4}A_{331})
+{\overline{{\bf{g}}^{*}}}_x{\overline{{\bf{g}}^{*}}}_y(\frac{3}{4}A_{312}+\frac{3}{4}A_{322}+\frac{1}{4}A_{332})
+{\overline{{\bf{g}}^{*}}}_x{\overline{{\bf{g}}^{*}}}_z(\frac{3}{4}A_{313}+\frac{1}{4}A_{323}+\frac{3}{4}A_{333})\notag\\
\mathcal{H}_{1y}&
={\overline{{\bf{g}}^{*}}}_y{\overline{{\bf{g}}^{*}}}_x(\frac{3}{4}A_{311}+\frac{3}{4}A_{321}+\frac{1}{4}A_{331})
+{\overline{{\bf{g}}^{*}}}_y^2(\frac{1}{4}A_{312}+\frac{3}{2}A_{322}+\frac{1}{4}A_{332})
+{\overline{{\bf{g}}^{*}}}_y{\overline{{\bf{g}}^{*}}}_z(\frac{1}{4}A_{313}+\frac{3}{4}A_{323}+\frac{3}{4}A_{333})\notag\\
\mathcal{H}_{1z}&
={\overline{{\bf{g}}^{*}}}_z{\overline{{\bf{g}}^{*}}}_x(\frac{3}{4}A_{311}+\frac{1}{4}A_{321}+\frac{3}{4}A_{331})
+{\overline{{\bf{g}}^{*}}}_z{\overline{{\bf{g}}^{*}}}_y(\frac{1}{4}A_{312}+\frac{3}{4}A_{322}+\frac{3}{4}A_{332})
+{\overline{{\bf{g}}^{*}}}_z^2(\frac{1}{4}A_{313}+\frac{1}{4}A_{323}+\frac{3}{2}A_{333})\notag\\
{\bf{a}}_{61}&
=\frac{1}{4}H_7\frac{1}{{\bar{g^*}}^4}({\bf{I}}:{\overline{{\bf{g}}^{*}}}\,{\overline{{\bf{g}}^{*}}})
({\overline{{\bf{g}}^{*}}}\,{\overline{{\bf{g}}^{*}}}\cdot{\bf{A}}_6)+\frac{5}{4}H_8{\bf{A}}_6
+\frac{7}{4}\frac{H_9}{{\bar{g^*}}^2}{\overline{{\bf{g}}^{*}}}\,{\overline{{\bf{g}}^{*}}}\cdot{\bf{A}}_6
+\frac{1}{4}\frac{H_9}{{\bar{g^*}}^2}({\overline{{\bf{g}}^{*}}}\,{\overline{{\bf{g}}^{*}}}:{\bf{I}}){\bf{I}}\cdot{\bf{A}}_6\notag\\
{\bf{a}}_{62}&
=\frac{1}{4}H_7\frac{1}{{\bar{g^*}}^4}{\overline{{\bf{g}}^{*}}}[\overline{{\bf{G}}^{\#}}({\overline{{\bf{g}}^{*}}}
\cdot{{\bm{\mathcal{W}}}_2}):{\overline{{\bf{g}}^{*}}}\,{\overline{{\bf{g}}^{*}}}]
+(\bar{G}^{\#}_{sx}+\bar{G}^{\#}_{sy}+\bar{G}^{\#}_{sz})H_9\frac{1}{{\bar{g^*}}^2}{\bm{\mathcal{H}}}_2\notag\\
+&(\bar{G}^{\#}_{sx}+\bar{G}^{\#}_{sy}+\bar{G}^{\#}_{sz})\frac{1}{4}H_8(3\mathcal{W}_{211}+\mathcal{W}_{221}+\mathcal{W}_{231},
\mathcal{W}_{212}+3\mathcal{W}_{222}+\mathcal{W}_{232},\mathcal{W}_{213}+\mathcal{W}_{223}+3\mathcal{W}_{233})\notag\\
{\bm{\mathcal{W}}}_2&\equiv\overline{{\bf{G}}^{\#}}{\bf{A}}_6\notag\\
{\bm{\mathcal{H}}}_2&\equiv(\mathcal{H}_{2x},\mathcal{H}_{2y},\mathcal{H}_{2z})\notag\\
\mathcal{H}_{2x}&
={\overline{{\bf{g}}^{*}}}_x^2(\frac{3}{2}\mathcal{W}_{211}+\frac{1}{4}\mathcal{W}_{221}+\frac{1}{4}\mathcal{W}_{231})
+{\overline{{\bf{g}}^{*}}}_x{\overline{{\bf{g}}^{*}}}_y(\frac{3}{4}\mathcal{W}_{212}+\frac{3}{4}\mathcal{W}_{222}+\frac{1}{4}\mathcal{W}_{223})
+{\overline{{\bf{g}}^{*}}}_x{\overline{{\bf{g}}^{*}}}_z(\frac{3}{4}\mathcal{W}_{213}+\frac{1}{4}\mathcal{W}_{223}+\frac{3}{4}\mathcal{W}_{333})\notag\\
\mathcal{H}_{2y}&
={\overline{{\bf{g}}^{*}}}_y{\overline{{\bf{g}}^{*}}}_x(\frac{3}{4}\mathcal{W}_{211}+\frac{3}{4}\mathcal{W}_{221}+\frac{1}{4}\mathcal{W}_{231})
+{\overline{{\bf{g}}^{*}}}_y^2(\frac{1}{4}\mathcal{W}_{212}+\frac{3}{2}\mathcal{W}_{222}+\frac{1}{4}\mathcal{W}_{223})
+{\overline{{\bf{g}}^{*}}}_y{\overline{{\bf{g}}^{*}}}_z(\frac{1}{4}\mathcal{W}_{213}+\frac{3}{4}\mathcal{W}_{223}+\frac{3}{4}\mathcal{W}_{333})\notag\\
\mathcal{H}_{2z}&
={\overline{{\bf{g}}^{*}}}_z{\overline{{\bf{g}}^{*}}}_x(\frac{3}{4}\mathcal{W}_{211}+\frac{1}{4}\mathcal{W}_{221}+\frac{3}{4}\mathcal{W}_{231})
+{\overline{{\bf{g}}^{*}}}_z{\overline{{\bf{g}}^{*}}}_y(\frac{1}{4}\mathcal{W}_{212}+\frac{3}{4}\mathcal{W}_{222}+\frac{3}{4}\mathcal{W}_{223})
+{\overline{{\bf{g}}^{*}}}_z^2(\frac{1}{4}\mathcal{W}_{213}+\frac{1}{4}\mathcal{W}_{223}+\frac{3}{2}\mathcal{W}_{333}).\notag\\
\label{a35_1}
\end{align}
}}

\subsubsection{Energy dissipation rate}
Energy dissipation rate is the vital term in the granular temperature equation. In our previous work \citep{zhao2021kinetic}, we have distinguished the contributions between non-local and inelastic collision by defining the perturbation functions by two parameters as the Knudsen number and $\epsilon\equiv(1-e_{ij}^2)$. But in this work, the perturbation function with the single parameter (the Knudsen number) has been selected, all effect of inelastic collision comes from the particle velocity variation between per- or post-collision, and the PVDF with the perturbation function has no relationship with the inelastic particle collision. Therefore, the energy dissipation rate of species $i$ (${{N}}_{di}={{N}}_{di}^{O(1)}+{{N}}_{di}^{O(K)}$) in the polydisperse system is the sum of following two parts:
\begin{align}
&{{N}}_{di}^{O(1)}\notag\\
=&\sum_j^N\bigg\{d_{ij}^2g_{ij}m_i\int\int\int\limits_{({\bf{g}}\cdot{\bf{k}})>0}
({\bf{c}}_i^{'2}-{\bf{c}}_i^2)f_i^{(0)}f_j^{(0)}({\bf{g}}\cdot{\bf{k}})d{\bf{k}}d{\bf{c}}_id{\bf{c}}_j\notag\\
+&\frac{d_{ij}^3}{2}g_{ij}m_i\int\int\int\limits_{({\bf{g}}\cdot{\bf{k}})>0}
({\bf{c}}_i^{'2}-{\bf{c}}_i^2)f_i^{(0)}f_j^{(0)}\nabla\ln\frac{f_i^{(0)}}{f_j^{(0)}}\cdot{\bf{k}}({\bf{g}}\cdot{\bf{k}})
d{\bf{k}}d{\bf{c}}_id{\bf{c}}_j\bigg\}\notag\\
=&\sum_j^N\bigg\{\notag\\
&3\pi^{-\frac{1}{2}}g_{ij}\varepsilon_i\varepsilon_j\rho_i\frac{d_{ij}^2}{d_{j}^3}(\theta_i+\theta_j)
\bigg[\bigg(\eta_i^2-2\eta_i\frac{\theta_i}{\theta_i+\theta_j}\bigg)B^{-\frac{1}{2}}\mathcal{L}_3
+\bigg(2\eta_i\bigg)(\overline{{\bf{G}}^{\#}}\cdot\overline{{\bf{g}}^{*}})\mathcal{L}_2\bigg]
\notag\\
+&\frac{3\times2^{\frac{3}{2}}}{5}g_{ij}\varepsilon_i\varepsilon_j\rho_i\bigg(\frac{d_{ij}}{d_j}\bigg)^3
\bigg[\eta_i^2-2\eta_i\frac{\theta_i}{\theta_i+\theta_j}\bigg](\theta_i+\theta_j)^{\frac{3}{2}}
\notag\\
&\quad\times\bigg\{\bigg[\frac{1}{2}(5+2\overline{{\bf{g}}^{*}}^2){\bf{A}}_1
-\frac{1}{4}A^{-1}(3+2A\overline{{\bf{G}}^{\#}}^2)(5+2\overline{{\bf{g}}^{*}}^2){\bf{A}}_4
-\frac{1}{4}B^{-1}(35+2\overline{{\bf{g}}^{*}}^2(7+\overline{{\bf{g}}^{*}}^2)){\bf{A}}_5\bigg]\cdot\overline{{\bf{g}}^{*}}\notag\\
&\quad\quad\quad+\frac{1}{2}(5+2\overline{{\bf{g}}^{*}}^2)\overline{{\bf{g}}^{*}}\overline{{\bf{G}}^{\#}}:{\bf{A}}_2
-B^{-\frac{1}{2}}
\bigg[\frac{1}{4}(5+2\overline{{\bf{g}}^{*}}^2){\bf{I}}+\frac{1}{2}(7+2\overline{{\bf{g}}^{*}}^2)\overline{{\bf{g}}^{*}}\overline{{\bf{g}}^{*}}\bigg]
:[{\bf{A}}_3+{\bf{A}}_6\overline{{\bf{G}}^{\#}}]\bigg\}\notag\\
&+\frac{1}{5}g_{ij}\varepsilon_i\varepsilon_j\rho_i\bigg(\frac{d_{ij}}{d_j}\bigg)^3
\bigg[2\eta_i\bigg](\theta_i+\theta_j)\notag\\
&\quad\times\bigg\{[(5+2\overline{{\bf{g}}^{*}}^2){\bf{I}}+4\overline{{\bf{g}}^{*}}\overline{{\bf{g}}^{*}}]
:[\overline{{\bf{G}}^{\#}}{\bf{A}}_1+\frac{1}{2}A^{-1}{\bf{A}}_2+\overline{{\bf{G}}^{\#}}(\overline{{\bf{G}}^{\#}}\cdot{\bf{A}}_2)]\notag\\
&\quad\quad\quad
-\bigg[2({\bf{I}}:{\bf{A}}_3)(\overline{{\bf{g}}}\cdot\overline{{\bf{G}}^{\#}})
+(9+2\overline{{\bf{g}}^{*}}^2)(\overline{{\bf{G}}^{\#}}\overline{{\bf{g}}}:{\bf{A}}_3)
+4B(\bar{\bf{g}}\bar{\bf{g}}\cdot\overline{{\bf{G}}^{\#}})\cdot(\bar{\bf{g}}\cdot{\bf{A}}_3)
\bigg]\notag\\
&\quad\quad\quad
-A^{-1}(5+2A\overline{{\bf{G}}^{\#}}^2)(\frac{5}{2}{\bf{I}}+2\overline{{\bf{g}}^{*}}\,\overline{{\bf{g}}^{*}})
:\overline{{\bf{G}}^{\#}}{\bf{A}}_4\notag\\
&\quad\quad\quad
-B^{-1}\bigg[(8+4\overline{{\bf{g}}^{*}}^2){\bf{I}}
+2(7+2\overline{{\bf{g}}^{*}}^2)\overline{{\bf{g}}^{*}}\,\overline{{\bf{g}}^{*}}\bigg]:\overline{{\bf{G}}^{\#}}{\bf{A}}_5\notag\\
&\quad\quad\quad
-A^{-1}B^{-\frac{1}{2}}\bigg[(5+2\overline{{\bf{g}}^{*}}^2)[(\frac{1}{2}{\bf{I}}+A\overline{{\bf{G}}^{\#}}\,\overline{{\bf{G}}^{\#}})\cdot\overline{{\bf{g}}^{*}}]\cdot{\bf{A}}_6
+AB^{\frac{1}{2}}[4{\bf{A}}_6\cdot(\bar{\bf{g}}\cdot\overline{{\bf{G}}^{\#}}\,\overline{{\bf{G}}^{\#}})
+2(\bar{\bf{g}}\cdot{\bf{A}}_6)\bigg(\overline{{\bf{G}}^{\#}}\,\overline{{\bf{G}}^{\#}}:({\bf{I}}+2B\bar{\bf{g}}\bar{\bf{g}})\bigg)]
\bigg]\bigg\}\notag\\
&\bigg\},
\end{align}

\begin{align}
&N_{di}^{O(K)}\notag\\
=&\sum_j^{N}\bigg\{\frac{d_{ij}^2}{2}m_ig_{ij}\int({\bf{c}}_i^{'2}-{\bf{c}}_i^2)f_i^{(0)}f_j^{(0)}(\Phi_i^{(K)}+\Phi_j^{(K)})({\bf{g}}\cdot{\bf{k}})d{\bf{k}}d{\bf{c}}_id{\bf{c}}_j\bigg\}\notag\\
=&\sum_j^{N}\bigg\{\notag\\
&\frac{d_{ij}^2}{2}m_ig_{ij}\bigg(\frac{1}{4\pi^2\theta_i\theta_j}\bigg)^{\frac{3}{2}}n_in_j\bigg\{\bigg(\eta_i^2-2\eta_i\frac{\theta_i}{\theta_i+\theta_j}
\bigg)\frac{\pi}{2}
\bigg[-A_iB_i(\pi^{\frac{3}{2}}A^{-\frac{3}{2}}B^{-3})
\bigg(\pi\mathcal{L}_3(\frac{5}{4\theta_i}A^{-1}{\bf{R}}_i-\frac{5}{2}{\bf{R}}_i+\frac{1}{2\theta_i}{\bf{R}}_i^2{\bf{R}}_i)\notag\\
&+B^{-\frac{1}{2}}H_6(\overline{{{g}}^{*}})\exp(-\overline{{{g}}^{*}}^2)\frac{\theta_i}{\theta_i+\theta_j}
(-\frac{5}{4\theta_i}A^{-1}\overline{{\bf{g}}^{*}}+\frac{5}{2}\overline{{\bf{g}}^{*}}-\frac{1}{2\theta_i}{\bf{R}}_i^2\overline{{\bf{g}}^{*}}
-\frac{1}{\theta_i}\overline{{\bf{g}}^{*}}\cdot{\bf{R}}_i{\bf{R}}_i)\notag\\
&+B^{-1}\exp(-\overline{{{g}}^{*}}^2)\frac{\theta_i}{(\theta_i+\theta_j)^2}
(\frac{1}{2}{\bf{R}}_i\cdot(\mathcal{H}_9{\bf{I}}+H_6{\bf{I}}+\mathcal{H}_8\overline{{\bf{g}}^{*}}\overline{{\bf{g}}^{*}})
-\frac{\theta_i}{2(\theta_i+\theta_j)}B^{-\frac{1}{2}}\mathcal{L}_4\overline{{\bf{g}}^{*}})\bigg)\cdot\nabla\ln\theta_i\notag\\
&-A_iD_i(\pi^{\frac{3}{2}}A^{-\frac{3}{2}}B^{-3})\bigg(\frac{\theta_i^2}{(\theta_i+\theta_j)^2}B^{-1}\exp(-\overline{{{g}}^{*}}^2)
(-\frac{1}{3}\mathcal{H}_9{\bf{I}}+\frac{1}{2}(H_6{\bf{I}}+\mathcal{H}_8\overline{{\bf{g}}^{*}}\overline{{\bf{g}}^{*}}))\notag\\
&-\frac{2\theta_i}{\theta_i+\theta_j}B^{-\frac{1}{2}}H_6\exp(-\overline{{{g}}^{*}}^2)
[-\frac{1}{3}({\bf{R}}_i\cdot\overline{{\bf{g}}^{*}}){\bf{I}}+\frac{1}{2}({\bf{R}}_i\overline{{\bf{g}}^{*}}+\overline{{\bf{g}}^{*}}{\bf{R}}_i)]+\pi\mathcal{L}_3({\bf{R}}_i{\bf{R}}_i-\frac{1}{3}{{R}}_i^2{\bf{I}})\bigg):\nabla{\bf{u}}_i\notag\\
&-A_jB_j(\pi^{\frac{3}{2}}A^{-\frac{3}{2}}B^{-3})
\bigg(\pi\mathcal{L}_3(\frac{5}{4\theta_j}A^{-1}{\bf{R}}_j-\frac{5}{2}{\bf{R}}_j+\frac{1}{2\theta_j}{\bf{R}}_j^2{\bf{R}}_j)\notag\\
&+B^{-\frac{1}{2}}H_6\exp(-\overline{{{g}}^{*}}^2)\frac{\theta_j}{\theta_i+\theta_j}
(\frac{5}{4\theta_j}A^{-1}\overline{{\bf{g}}^{*}}-\frac{5}{2}\overline{{\bf{g}}^{*}}+\frac{1}{2\theta_j}{\bf{R}}_j^2\overline{{\bf{g}}^{*}}
+\frac{1}{\theta_i}\overline{{\bf{g}}^{*}}\cdot{\bf{R}}_j{\bf{R}}_j)\notag\\
&+B^{-1}\exp(-\overline{{{g}}^{*}}^2)\frac{\theta_j}{(\theta_i+\theta_j)^2}
[\frac{1}{2}{\bf{R}}_j\cdot(\mathcal{H}_9{\bf{I}}+H_6{\bf{I}}+\mathcal{H}_8\overline{{\bf{g}}^{*}}\overline{{\bf{g}}^{*}})
+\frac{\theta_j}{2(\theta_i+\theta_j)}B^{-\frac{1}{2}}\mathcal{L}_4\overline{{\bf{g}}^{*}}]\bigg)\cdot\nabla\ln\theta_j\notag\\
&-A_jD_j(\pi^{\frac{3}{2}}A^{-\frac{3}{2}}B^{-3})
\bigg(\frac{2\theta_j}{\theta_i+\theta_j}B^{-\frac{1}{2}}H_6\exp(-\overline{{{g}}^{*}}^2)(\frac{1}{2}({\bf{R}}_j\overline{{\bf{g}}^{*}}+\overline{{\bf{g}}^{*}}{\bf{R}}_j)-\frac{1}{3}({\bf{R}}_j\cdot\overline{{\bf{g}}^{*}}){\bf{I}})
\notag\\
&+\frac{\theta_j^2}{(\theta_i+\theta_j)^2}B^{-1}\exp(-\overline{{{g}}^{*}}^2)
(\frac{1}{2}(H_6{\bf{I}}+\mathcal{H}_8\overline{{\bf{g}}^{*}}\overline{{\bf{g}}^{*}})-\frac{1}{3}\mathcal{H}_9{\bf{I}})
+\pi\mathcal{L}_3({\bf{R}}_j{\bf{R}}_j-\frac{1}{3}{\bf{R}}_j^2{\bf{I}})\bigg):\nabla{\bf{u}}_j\bigg]\notag\\
&+\bigg(2\eta_i\bigg)\frac{\pi}{2}\bigg[-A_iB_i(\pi^{\frac{3}{2}}A^{-\frac{5}{2}}B^{-\frac{5}{2}})
\bigg(h_3\exp(-\overline{{{g}}^{*}}^2)(\frac{5}{8\theta_i}A^{-1}\overline{{\bf{g}}^{*}}+\frac{1}{4\theta_i}{\bf{R}}_i^2\overline{{\bf{g}}^{*}}-\frac{5}{4}\overline{{\bf{g}}^{*}})\notag\\
&+B^{-1}\bigg(\frac{3\theta_i}{4(\theta_i+\theta_j)^2}\overline{{\bf{g}}^{*}}\bigg)H_6\exp(-\overline{{{g}}^{*}}^2)
-\frac{B^{-\frac{1}{2}}\pi^{\frac{1}{2}}}{2(\theta_i+\theta_j)}
[(h_6{\bf{I}}+h_7\frac{\overline{{\bf{g}}^{*}}\,\overline{{\bf{g}}^{*}}}{\overline{{\bf{g}}^{*}}^2})\cdot{\bf{R}}_i
+2{\bf{I}}:(h_6{\bf{I}}+h_7\frac{\overline{{\bf{g}}^{*}}\,\overline{{\bf{g}}^{*}}}{\overline{{\bf{g}}^{*}}^2}){\bf{R}}_i]\notag\\
&+[h_3\exp(-\overline{{{g}}^{*}}^2)(\frac{5}{4\theta_i}+\frac{1}{2\theta_i}A{\bf{R}}_i^2-\frac{5}{2}A)
+H_6\frac{\theta_i}{2(\theta_i+\theta_j)^2}]({\bf{R}}_i\overline{{\bf{G}}^{\#}}\cdot\overline{{\bf{g}}^{*}})\notag\\
&+[\pi B^{-\frac{1}{2}}\exp(-\overline{{{g}}^{*}}^2)(h_6{\bf{I}}+h_7\frac{\overline{{\bf{g}}^{*}}\,\overline{{\bf{g}}^{*}}}{\overline{{\bf{g}}^{*}}^2})
(-\frac{5}{4}\frac{1}{\theta_i+\theta_j}-\frac{1}{2(\theta_i+\theta_j)}A{\bf{R}}_i^2
+\frac{5}{2}A\frac{\theta_i}{\theta_i+\theta_j})
-AB^{-\frac{3}{2}}(H_6{\bf{I}}+\mathcal{H}_8\overline{{\bf{g}}^{*}}\,\overline{{\bf{g}}^{*}})\frac{\theta_i^2}{4(\theta_i+\theta_j)^3}]\overline{{\bf{G}}^{\#}}\notag\\
&+AB^{-1}\frac{\theta_i}{(\theta_i+\theta_j)^2}\exp(-\overline{{{g}}^{*}}^2)
[h_4\overline{{\bf{g}}^{*}}(\overline{{\bf{g}}^{*}}\,\overline{{\bf{g}}^{*}}:\overline{{\bf{G}}^{\#}}{\bf{R}}_i)
+\pi h_5(\overline{{\bf{g}}^{*}}\cdot(\overline{{\bf{G}}^{\#}}{\bf{R}}_i+{\bf{R}}_i\overline{{\bf{G}}^{\#}}
+(\overline{{\bf{G}}^{\#}}\cdot{\bf{R}}_i){\bf{I}}))]\notag\\
&-\pi A B^{-\frac{1}{2}}(h_6{\bf{I}}+h_7\frac{\overline{{\bf{g}}^{*}}\,\overline{{\bf{g}}^{*}}}{\overline{{\bf{g}}^{*}}^2})
:(\frac{1}{\theta_i+\theta_j}\overline{{\bf{G}}^{\#}}{\bf{R}}_i){\bf{R}}_i\bigg)\cdot\nabla\ln\theta_i\notag\\
&-A_iD_i(\pi^{\frac{3}{2}}A^{-\frac{3}{2}}B^{-\frac{5}{2}})
\bigg(A^{-1}h_3\exp(-\overline{{{g}}^{*}}^2)(1-\frac{\theta_i}{\theta_i+\theta_j})[\frac{1}{2}(\overline{{\bf{g}}^{*}}{\bf{R}}_i+{\bf{R}}_i\overline{{\bf{g}}^{*}})+({\bf{R}}_i\cdot\overline{{\bf{g}}^{*}}){\bf{I}}]\notag\\
&+h_3\exp(-\overline{{{g}}^{*}}^2)({\bf{R}}_i{\bf{R}}_i-\frac{1}{3}{\bf{R}}_i^2{\bf{I}})(\overline{{\bf{G}}^{\#}}\cdot\overline{{\bf{g}}^{*}})\notag\\
&+B^{-1}\frac{\theta_i^2}{(\theta_i+\theta_j)^2}
[h_4(\overline{{\bf{G}}^{\#}}\cdot\overline{{\bf{g}}^{*}}\overline{{\bf{g}}^{*}})\overline{{\bf{g}}^{*}}
+\pi h_5((\overline{{\bf{G}}^{\#}}\cdot\overline{{\bf{g}}^{*}}){\bf{I}}+(\overline{{\bf{G}}^{\#}}\overline{{\bf{g}}^{*}}
+\overline{{\bf{g}}^{*}}\overline{{\bf{G}}^{\#}}))-\frac{1}{3}H_6\exp(-\overline{{{g}}^{*}}^2)\overline{{\bf{G}}^{\#}}\overline{{\bf{g}}^{*}}]\notag\\
&-\pi B^{-\frac{1}{2}}
\frac{2\theta_i}{\theta_i+\theta_j}\exp(-\overline{{{g}}^{*}}^2)
[h_6(\frac{1}{2}(\overline{{\bf{G}}^{\#}}{\bf{R}}_i+{\bf{R}}_i\overline{{\bf{G}}^{\#}})
+(\overline{{\bf{G}}^{\#}}\cdot{\bf{R}}_i){\bf{I}})
+h_7\frac{1}{\overline{{{g}}^{*}}^2}
(\frac{1}{2}(\overline{{\bf{G}}^{\#}}\cdot\overline{{\bf{g}}^{*}})({\bf{R}}_i\overline{{\bf{g}}^{*}}+\overline{{\bf{g}}^{*}}{\bf{R}}_i)
+(\overline{{\bf{G}}^{\#}}{\bf{R}}_i:\overline{{\bf{g}}^{*}}\overline{{\bf{g}}^{*}}){\bf{I}})]
\bigg):\nabla{\bf{u}}_i,\notag\\
\label{a36}
\end{align}
{{where the forms of $\mathcal{L}_2$,$\mathcal{L}_3$ and $\mathcal{L}_4$ are:
\begin{equation}
\begin{split}
&{\mathcal{L}}_2(\bar{g}^{*})=(1+\frac{1}{2\bar{{g}}^{*2}})\exp(-\bar{{g}}^{*2})
+\sqrt{\pi}(\frac{1}{\bar{g}^{*}}+\bar{g}^{*}
-\frac{1}{4\bar{g}^{*3}})erf(\bar{g}^{*})
\\
&{\mathcal{L}}_3(\bar{g}^{*})
=\frac{1}{4\bar{g}^{*}}
\bigg\{2\bar{g}^{*}(5+2\bar{g}^{*2})\exp(-\bar{g}^{*2})
+\sqrt{\pi}
[3+4\bar{g}^{*2}(3+\bar{g}^{*2})]erf(\bar{g}^{*})\bigg\}\\
&\mathcal{L}_4
=\frac{\pi}{16\overline{{{g}}^{*}}^2}\bigg[2\overline{{{g}}^{*}}(5+2\overline{{{g}}^{*}}^2)
(3+4\overline{{{g}}^{*}}^2(7+\overline{{{g}}^{*}}^2))
+\exp(\overline{{{g}}^{*}}^2)\sqrt{\pi}
(-15+8\overline{{{g}}^{*}}^2(15+45\overline{{{g}}^{*}}^2+20\overline{{{g}}^{*}}^4+2\overline{{{g}}^{*}}^6))erf(\overline{{{g}}^{*}})\bigg].
\label{x5_4}
\end{split}
\end{equation}
}}
{{
\section{Model validation: Simple shear flow}
\subsection{Mathematical model}
The analysis of hydrodynamic equations and constitutive relations for one-dimensional simple shear flow of bidisperse particles are following our previous work \citep{shi2022critical}. In this case, only the granular temperature equation is needed:
\begin{equation}
\begin{split}
-\mu_i\bigg(\frac{du_i}{dy}\bigg)^2=\mathcal{N}_{d,i}.
\end{split}
\label{a37}
\end{equation}
where the shear viscosity $\mu_i$ and the energy dissipation rate $\mathcal{N}_{d,i}$ of species $i$ of this model are
\begin{equation}
\begin{split}
\mu_i=&\frac{5\sqrt{\pi}}{96}\frac{d_i\rho_i}{g_{ii}}\bigg(1+\frac{8}{5}g_{ii}\varepsilon_i\bigg)\theta_i^{1/2}\\
&+\sum_j^2\Bigg\{\frac{2\pi}{15}(1+e_{ij})d_{ij}^3g_{ij}\frac{\varepsilon_i\varepsilon_j\rho_i\rho_j}{m_0}
\bigg[\frac{5\sqrt{\pi}}{96}\frac{d_i}{g_{ii}\varepsilon_i}\big(1+\frac{8}{5}g_{ii}\varepsilon_i\big)\theta_i^{1/2}
+\frac{5\sqrt{\pi}}{96}\frac{d_j}{g_{jj}\varepsilon_j}\big(1+\frac{8}{5}g_{jj}\varepsilon_j\big)\theta_j^{1/2}\bigg]\Bigg\}\\
&+\sum_j^2\Bigg\{\frac{2\sqrt{2\pi}}{15}(1+e_{ij})d_{ij}^4g_{ij}\frac{\varepsilon_i\varepsilon_j\rho_i\rho_j}{m_0}(\theta_i+\theta_j)^{1/2}\Bigg\}.
\end{split}
\label{a38}
\end{equation}
\begin{equation}
\begin{split}
\mathcal{N}_{d,i}=&\sum_j^2\bigg\{2\sqrt{2\pi}{d^2_{ij}}g_{ij}\frac{\varepsilon_i\varepsilon_j\rho_i\rho_j}{m_0}(\theta_i+\theta_j)^{3/2}
\bigg[(1+e_{ij})^2\frac{m_j}{m_0}-2(1+e_{ij})\frac{\theta_i}{\theta_i+\theta_j}\bigg]\bigg\}.\\
\label{a39}
\end{split}
\end{equation}
For comparing the refined model with the pervious one of \citep{iddir2005modeling}, the corresponding shear viscosity $\mu_{i,Iddir}$ and  the energy dissipation rate $\mathcal{N}_{d,i,Iddir}$ deduced by simplifying the models of \cite{iddir2005modeling} are outlined as:
\begin{equation}
\begin{split}
{{\mu_{i,Iddir}}}
=&\underset{\textcircled{1}}{\uline{\frac{5\sqrt{\pi}}{96}\frac{d_i\rho_i}{g_{ii}}\bigg(1+\frac{4(1+e_{ii})}{5}g_{ii}\varepsilon_i\bigg)\bigg(\frac{T_i}{m_i}\bigg)^{1/2}}}\\
&\underset{\textcircled{2}}{\uline{+\frac{4\pi}{15}d_{i}^3g_{ii}\frac{\varepsilon_i\varepsilon_i\rho_i\rho_i}{2m_i}
\bigg[\frac{5(1+e_{ii})\sqrt{\pi}}{96}\frac{d_i}{g_{ii}\varepsilon_i}\big(1+\frac{4(1+e_{ii})}{5}g_{ii}\varepsilon_i\big)\bigg(\frac{T_i}{m_i}\bigg)^{1/2}\bigg]}}\\
&\underset{\textcircled{3}}{\uline{+\sum\limits^2_{j=1}\bigg\{\frac{2\sqrt{2\pi}}{15}(1+e_{ij})d^4_{ij}g_{ij}\frac{\rho_i\rho_j\varepsilon_j\varepsilon_i}{m_0}\bigg(\frac{T_i}{m_i}+\frac{T_j}{m_j}\bigg)^{1/2}
\frac{1}{2}\bigg(\frac{T_i}{m_i}+\frac{T_j}{m_j}\bigg)^{-2}(T_i+T_j)m_0^5\bigg(\frac{T_i}{m_i}\frac{T_j}{m_j}\bigg)^2(m_jT_j+m_iT_i)^{-3}(1+9B_{ij}^2A_{ij}^{-1}D_{ij}^{-1})\bigg\}}}\\
\end{split}
\label{a39_1}
\end{equation}
\begin{equation}
\begin{split}
{\mathcal{N}_{d,i,Iddir}}= &\sum\limits^2_{j=1}\Bigg\{\frac{3}{4}d^2_{ij}(e_{ij}+1)g_{ij}\varepsilon_i\varepsilon_j\rho_i\rho_j\bigg(\frac{{m_i}{m_j}}{{T _i}{T _j}}\bigg)^{\frac{3}{2}}\frac{\sqrt{\pi}}{m_0}\bigg[B_{ij}\cdot{R_5}+(e_{ij}-1)\frac{m_j}{m_0}\frac{1}{6}{\cdot}R_1\bigg]\Bigg\},\\
\end{split}
\label{a39_2}
\end{equation}
It's noted that: (I) The first term of ${{\mu_{i,Iddir}}}$ denoted as $\textcircled{1}$ represents the kinetic part related to the perturbation function $\Phi_{i,Iddir}^{(K)}$. Its form differs from the corresponding term in our model due to the different coefficient $D_i$ in perturbation function $\Phi_{i,Iddir}^{(K)}$. Specifically, in their model, $D_{i,Iddir}$ directly introduces the restitutive coefficient into the perturbation function, while in our model, $D_{i}$ holds the same form as gases without the inelastic collision effect; (II) The second term of  ${{\mu_{i,Iddir}}}$
denoted as $\textcircled{2}$ is the collisional contribution related to the perturbation function of $\Phi_{i,Iddir}^{(K)}$, in addition to the difference in coefficient $D_i$, only the contribution from the internal-species particle interaction has been involved and the effect of the interspecies particle interaction has been neglected; (III) The third term of ${{\mu_{i,Iddir}}}$
denoted as $\textcircled{3}$ also  is the collisional contribution, and it is related to the equilibrium PVDF with the complex form as $f_i^{(0)}f_j^{(0)}\nabla\ln\frac{f_i^{(0)}}{f_j^{(0)}}$. And the energy dissipation rate ${\mathcal{N}_{d,i,Iddir}}$ also results from the contribution of the complex collisional integrals related to the Maxwellian PVDF.  Moreover, the main reason of these two complex forms of $\textcircled{3}$  and ${\mathcal{N}_{d,i,Iddir}}$ is that the Taylor series expansions have been utilized to these collisional integrals.

The associated radial distribution functions, which reflect the impact of solid concentration on the likelihood of particle interactions, are chosen as the identical forms as in \cite{iddir2005modeling}:
\begin{equation}
\begin{split}
g_{ij}=\frac{{d_jg_{ii}}+d_ig_{jj}}{2d_{ij}}, \ \
g_{ii}=\frac{1}{[1-(\varepsilon_i+\varepsilon_j)/{\varepsilon_{max}}]}+\frac{3d_i}{2}\sum\limits^2_{j=1}\frac{\varepsilon_j}{d_j},
\end{split}
\label{a40}
\end{equation}
where $\varepsilon_{max}=0.64$ is the maximum value of solid volume fraction. In above  equations, the normal restitution coefficient is $e_{ij}\equiv\frac{1}{2}(e_i+e_j)$, the arithmetic mean diameter of bidisperse particles is $d_{ij}\equiv\frac{1}{2}(d_i+d_j)$ and the sum of particle mass is $m_0\equiv m_i+m_j$.
The granular temperature of each species can then be calculated by solving the nonlinear equation system using the granular temperature equations.

For a better comparison with previous works, the total granular pressure ${\tau}_{normal}$ and the total shear stress $\tau_{shear}$ are defined as:
\begin{equation}
\begin{split}
\tau_{normal}=\sum\limits_{i=1}^2P_{k,i}+\sum\limits_{i=1}^2\sum\limits_{j=1}^2 {P_{c,ij}},
\end{split}
\label{a41}
\end{equation}
\begin{equation}
\begin{split}
\tau_{shear}=\sum\limits_{i=1}^2\bigg(\mu_i\frac{du_i}{dy}\bigg)=\frac{du}{dy}\sum\limits_{i=1}^2\mu_i.
\end{split}
\label{a42}
\end{equation}
It's noted that:
%(i) The solid stress tensor for species $i$ in equation (\ref{a26}) undergoes simplification to be more applicable for the simple shear bidisperse granular flow. The two components corresponding to different species are subsequently added and the resulting shear and normal stresses are presented in Eq.(\ref{a41}) and Eq.(\ref{a42});
(i) The strain rates of various species align with the specified shear rates of the bidisperse systems in discrete element simulation, that is, $\frac{du_i}{dy}\equiv\frac{du}{dy}$; (ii) The normal pressure comprises of the kinetic component $P_{k,i}$ and the particle collision component $P_{c,ij}$, which are respectively formulated as
\begin{equation}
\begin{split}
&P_{k,i}=\varepsilon_i\rho_i\theta_i,\\
&P_{c,ij}=\frac{\pi}{3}(1+e_{ij})d_{ij}^3g_{ij}\frac{\varepsilon_i\varepsilon_j\rho_i\rho_j}{m_0}(\theta_i+\theta_j).
\end{split}
\label{a43}
\end{equation}

\subsection{Simulation results}
The discrete element simulation results of one-dimensional shear flow containing bidisperse particles with Lees-Edwards periodic boundaries \citep{galvin2007hydrodynamic} are used to validate the constitutive relations, where the predictions of \cite{iddir2005modeling} are also presented for comparison.  The simulation settings and granular material properties have been elaborated in previous works \citep{galvin2007hydrodynamic,shi2022critical}.
\begin{figure}[H]
\centerline{\includegraphics[width=0.55\textwidth]{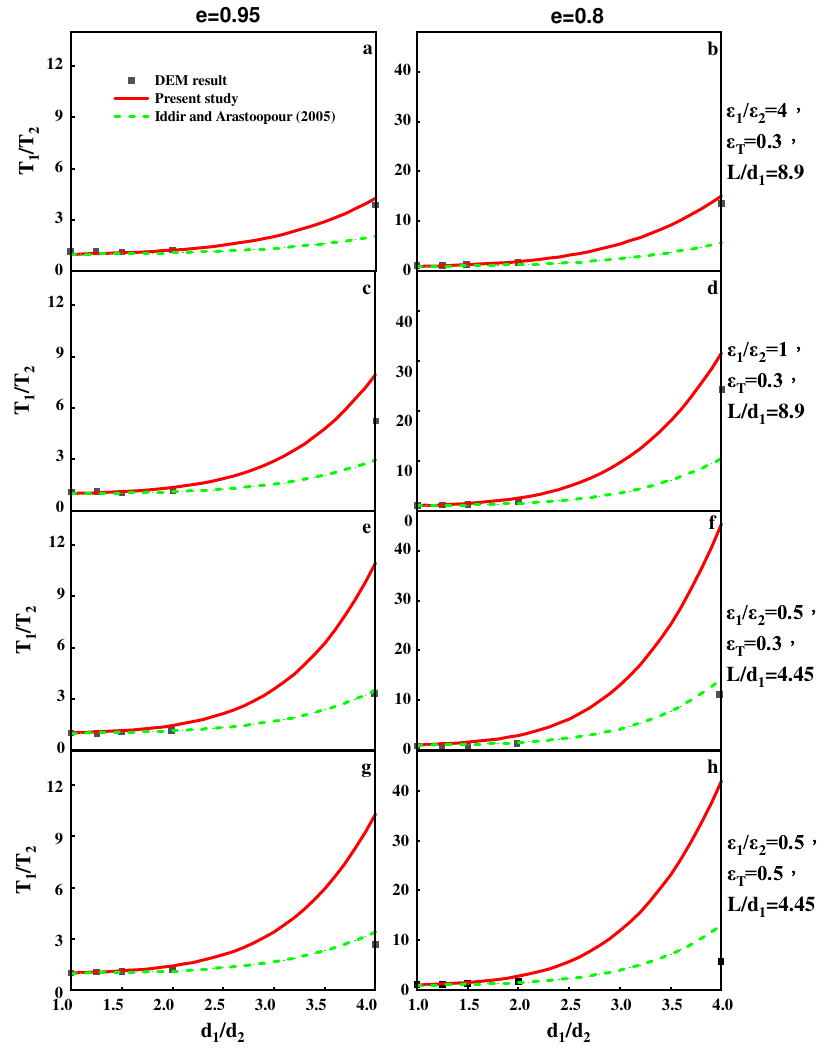}}
\caption{Granular temperature ratio $T_1/T_2$ as a function of the diameter ratio $d_1/d_2$ with $\rho_1/\rho_2=1$, for different restitution coefficients $e$, total solid volume fractions $\varepsilon_T$, ratio of solid volume fractions $\varepsilon_1/\varepsilon_2$ and ratio of the edge length of simulation domain to the diameter of bigger particle $L/d_1$.}\label{figs_1}
\end{figure}
Figure \ref{figs_1} shows that the variation of granular temperature ratio $T_1/T_2$ as a function of the diameter ratio $d_1/d_2$ at different total solid volume fractions $\varepsilon_T\equiv \varepsilon_1+\varepsilon_2$, ratio of the solid volume fractions $\varepsilon_1/\varepsilon_2$, restitution coefficients $e_{ij}\equiv e$, and ratio of the edge length of simulation domain to the diameter of bigger particles $L/d_1$. The present work and \cite{iddir2005modeling}
demonstrate reliable predictions of the granular temperature ratio in bidisperse shear granular flow, under most conditions.
However, as the simulation domain reduces and particle diameters of different species differ significantly, both methods lack the ability to predict the granular temperature ratio as Fig.(\ref{figs_1}e)-Fig.(\ref{figs_1}h).
Moreover, the predicted results by \cite{iddir2005modeling} are generally lower than from our present model as the diameter ratio $d_1/d_2$ increases.
Based on the comparison of shear viscosity and energy dissipation rate mentioned above, we have found that the shear viscosity of each species for binary shear granular flow is underestimated because (i) the collision part associated with $\Phi_i^{(K)}$, which reflects the interspecies particle interaction, has been neglected, and (ii) the part related to $f_i^{(0)}f_j^{(0)}\nabla\ln\frac{f_i^{(0)}}{f_j^{(0)}}$ at the first order was treated by the Taylor series expansion and only the initial terms were preserved. The energy dissipation rate resulting from the inelastic particle collision was calculated using the same Taylor series expansion approximation to account for the complex interspecies interaction. Based on these two underestimations, the single parameter C-E method proposed by \cite{iddir2005modeling} may predict a lower ratio of granular temperature.
\begin{figure}[H]
\centerline{\includegraphics[width=0.55\textwidth]{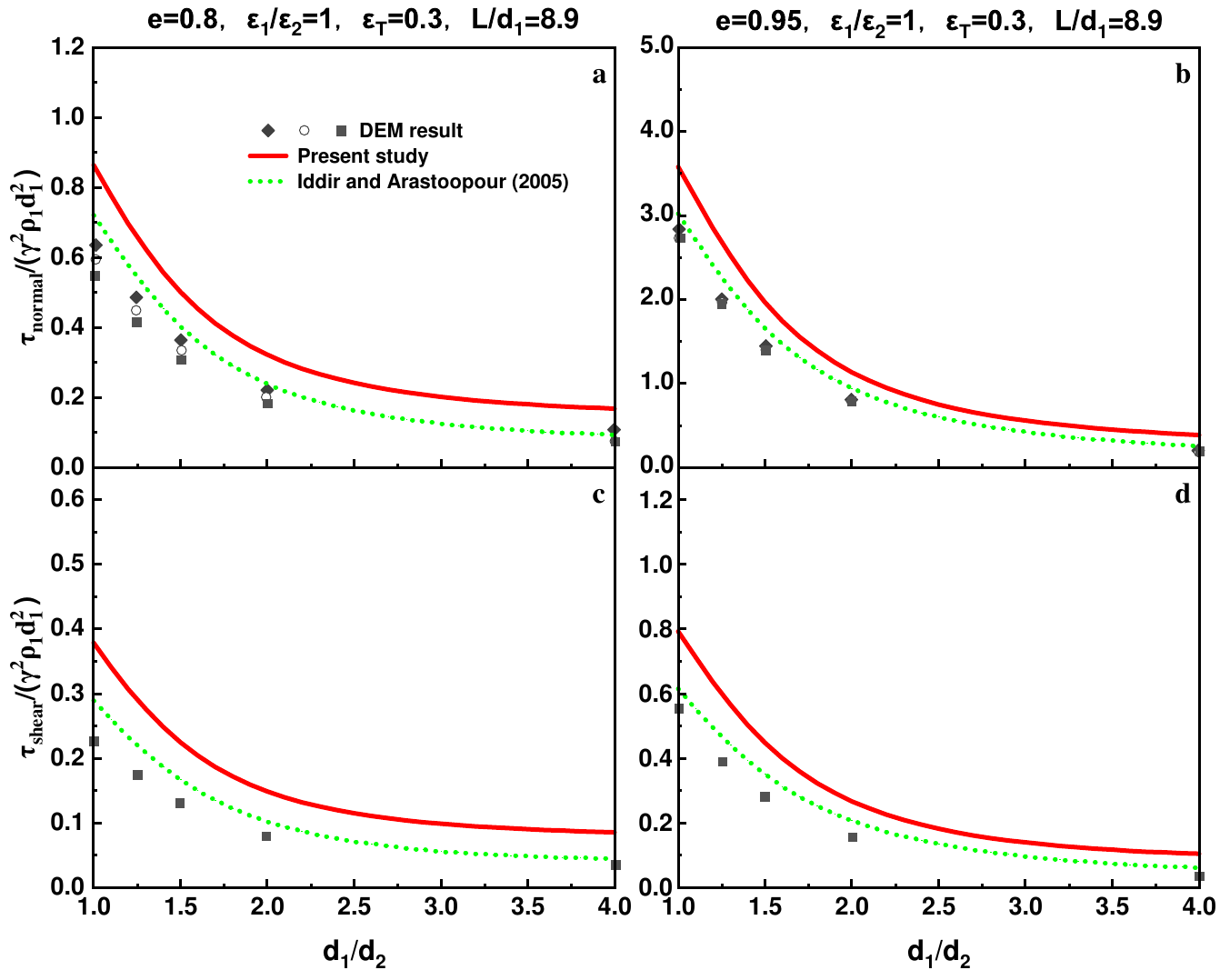}}
\caption{Non-dimensional normal and shear stresses $\tau_k/(\gamma^2\rho_1 d_1^2)$ plotted against the diameter ratio $d_1/d_2$ with $\rho_1/\rho_2=1$. In the non-dimensional normal stress, DEM simulations have three components, with $\tau_{xx}$ (diamonds), $\tau_{yy}$ (circles) and $\tau_{zz}$ (squares).}\label{figs_2}
\end{figure}

\begin{figure}[H]
\centerline{\includegraphics[width=0.55\textwidth]{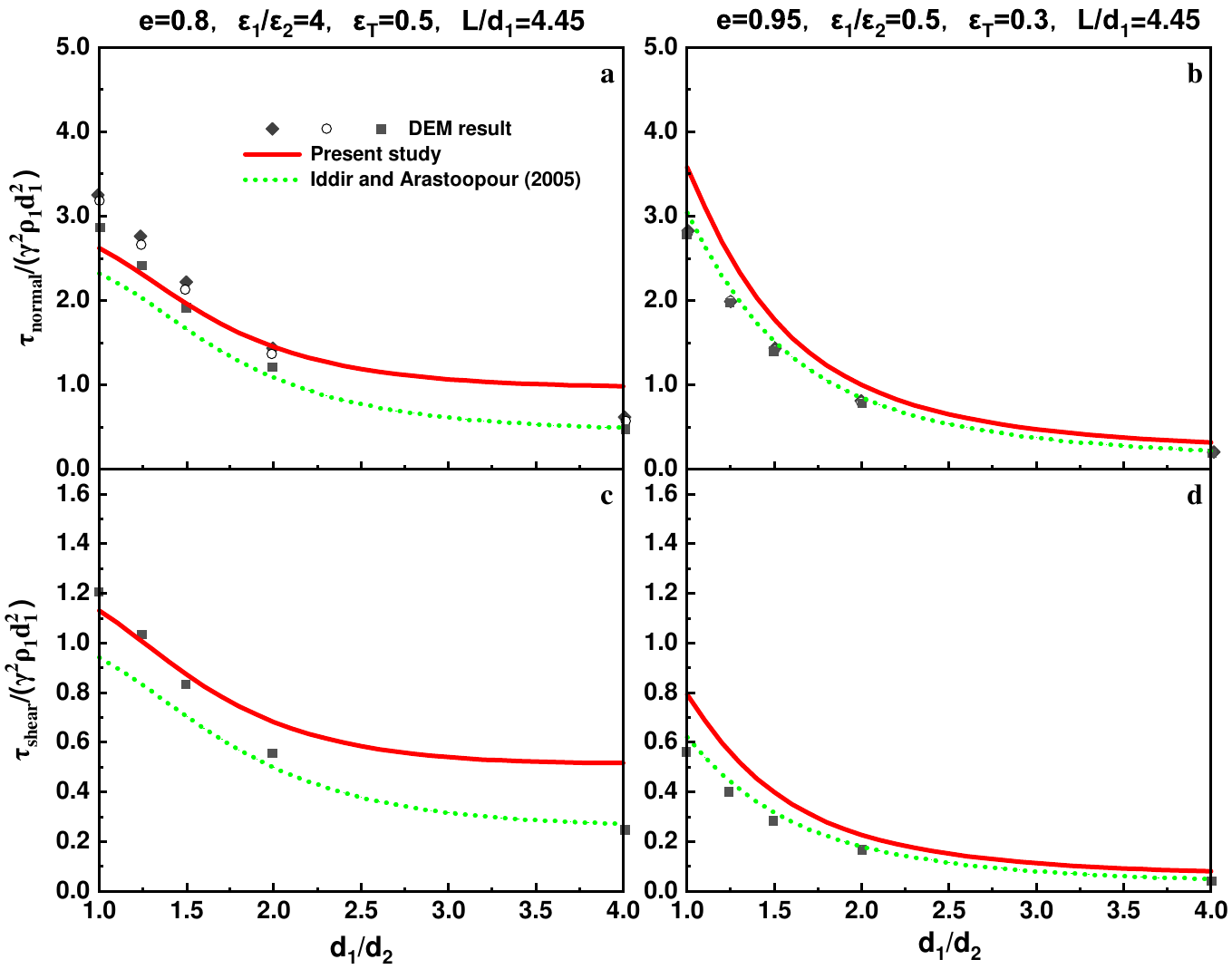}}
\caption{The same as Figure \ref{figs_2}, but now for different $\varepsilon_1/\varepsilon_2$, $\varepsilon_T$ and $L/d_1$.}\label{figs_3}
\end{figure}

\begin{figure}[H]
\centerline{\includegraphics[width=0.55\textwidth]{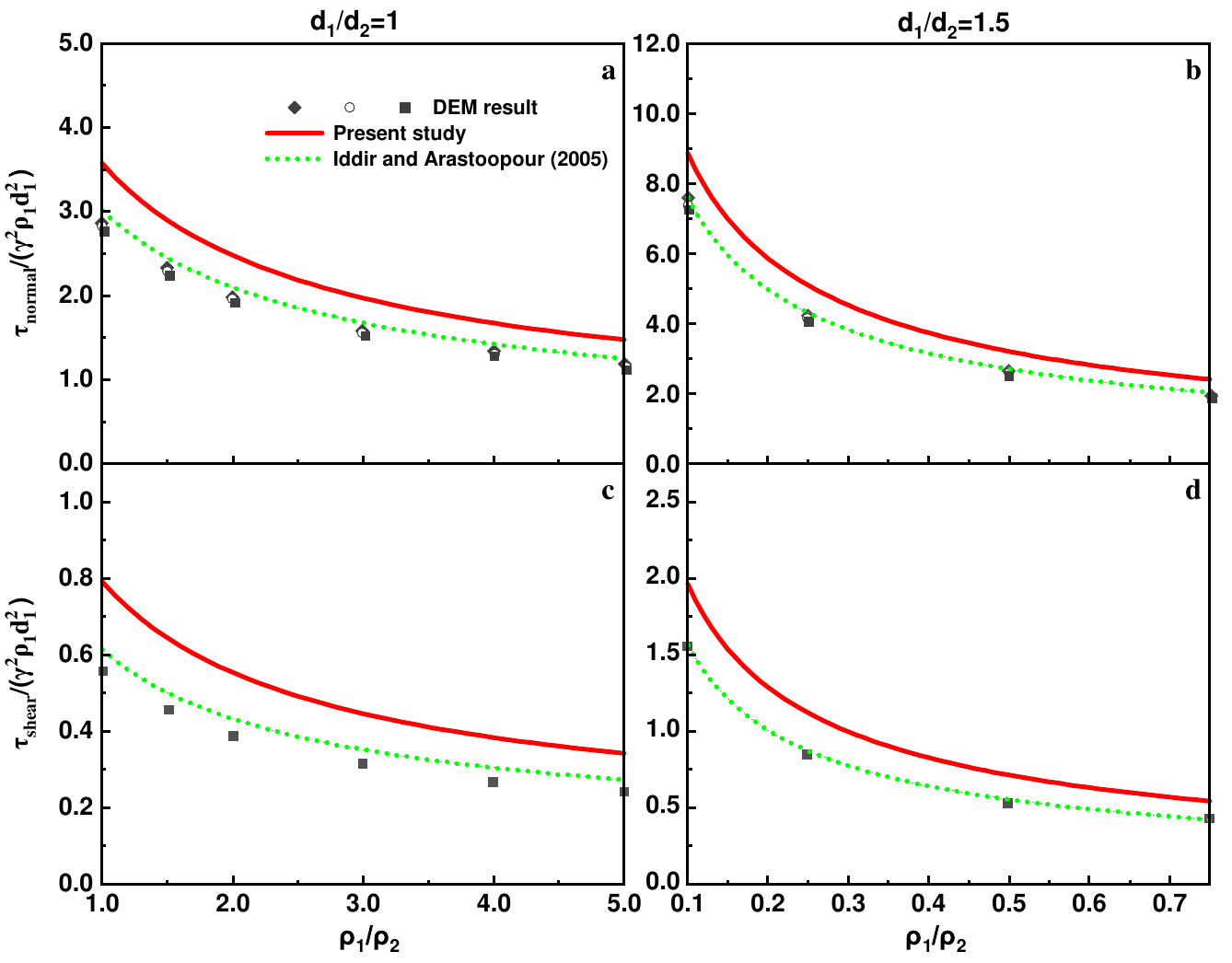}}
\caption{Non-dimensional normal and shear stresses $\tau_k/(\gamma^2\rho_1 d_1^2)$ plotted against the particle density ratio $\rho_1/\rho_2$.\\
In the non-dimensional normal stresses, DEM simulations with $\tau_{xx}$ (diamonds), $\tau_{yy}$ (circles) and $\tau_{zz}$ (squares).\\
Relevant parameters: e=0.95, $\varepsilon_1/\varepsilon_2=1$, $\varepsilon_T=0.3$, $L/d_1=8.9$.}\label{figs_4}
\end{figure}

Figure \ref{figs_2} and Figure \ref{figs_3} offer a comparison of the non-dimensional normal stress $\tau_{normal}/(\gamma^2\rho_1d_1^2)$ and shear stress $\tau_{shear}/(\gamma^2\rho_1d_1^2)$ as found though DEM simulations and kinetic theories with the particle diameter ratio with the equal mass density. And Figure {\ref{figs_4}} further reports the corresponding variations of with the particle density ratio at two different values of $d_1/d_2$.
It can be seen that although the present theory is a refinement of \cite{iddir2005modeling} in the sense that more rigorous mathematical derivations were carried out, the comparisons with these DEM data show that the present theory does not always outperform the theory of \cite{iddir2005modeling}.
As shown in Figure{\ref{figs_1}}, the predicted granular temperature ratio did not differ significantly between the present work and \cite{iddir2005modeling}. However, the dimensionless pressure and shear stress of the solid mixture as predicted by \cite{iddir2005modeling} are generally lower than the present results, which is consistent with the above analysis on the derivation of constitutive relations. We continue to try to modify the form of the perturbation function (\ref{a16}) following \cite{iddir2005modeling}. However, the corresponding revised results were inferior to theirs, indicating that using different perturbation function forms in the single-parameter C-E method does not yield different simulated results.  True associated reason made the different simulations can be shown as: Although our results are more accurate compared to those in \cite{iddir2005modeling} due to analytical and mathematical improvements, the associated perturbation functions do not account for the contribution from interspecies particle interaction in our work. This incomplete perturbation function may result in reduced normal pressure and shear stress of binary mixtures, and the reduced parts may be equal to the neglected high order parts in the Taylor series expansion introduced by \cite{iddir2005modeling}.

As shown in Figures ({\ref{figs_1}})-({\ref{figs_4}}), this refined KTGF has been directly validated using DEM data of one-dimensional shear flow. Furthermore, the refined model predicts mixture normal and shear stresses for bidisperse shear granular flow with weak inelasticity that are closer to the DEM simulations compared to \cite{iddir2005modeling}. In the future, the complete polydisperse C-E method needs to be developed, which must take into account the effect of interspecies interactions on the perturbation function.

}}

\section{Model validation: Bubbling fluidized bed}
\subsection{Mathematical model}
In this section, the multifluid model is utilized to simulate the particle mixing and segregation of binary gas-solid flow in a bubbling fluidized bed. The governing equations for binary gas-solid flow are outlined in
Tab.(\ref{tabA3}). The constitutive relations, including the solid stress tensor, the solid-solid drag force, the gas-solid drag force, the heat flux, and the energy dissipation rate, are summarized below.
{{It is important to note that the strategy that couples kinetic theory of granular flow with multifluid model for model validation is not always rigorous, since other models are needed, such as the gas-solid drag force and the frictional stress models.  However, since kinetic theory of polydisperse granular flow is a common selection for closing the particulate phase stresses of the multifluid model. Therefore, we confide that the validation work on bubbling fluidized bed is valuable.}}

\begin{table}[h]
\caption{The governing equations for binary gas-solid flow.}
\label{tabA3}
\begin{center}
\begin{tabular*}{0.9\hsize}{@{}@{\extracolsep{\fill}}llllll@{}}
 \hline
Mass conversation equations for gas and solid phase $i$:\\\\
$\frac{\partial}{\partial t}(\varepsilon_{g} \rho_{g})+\nabla \cdot(\varepsilon_{g} \rho_{g} \mathbf{u}_{g})=0$\\\\
$\frac{\partial}{\partial t}(\varepsilon_{i} \rho_{i})+\nabla \cdot(\varepsilon_{i} \rho_{i} \mathbf{u}_{i})=0$\\
\\
Momentum conservation equations for gas phase and solid phase $i$:\\\\
$\frac{\partial}{\partial t}(\varepsilon_{g} \rho_{g} \mathbf{u}_{g})+\nabla \cdot(\varepsilon_{g} \rho_{g} \mathbf{u}_{g} \mathbf{u}_{g})=-\varepsilon_g\nabla p_g+\nabla \cdot (\varepsilon_g{\bm{\tau}}_{g})+\varepsilon_{g} \rho_{g}{\dot{\bf{G}}}-\sum_{i=1}^2{\beta_{gi}({\bf{u}}_g-{\bf{u}}_i)}$\\\\
$\frac{\partial}{\partial t}(\varepsilon_{i} \rho_{i} \mathbf{u}_{i})+\nabla \cdot(\varepsilon_{i} \rho_{i} \mathbf{u}_{i} \mathbf{u}_{i})=-\varepsilon_i\nabla p_g-\nabla\cdot({\bf{P}}_i)+\varepsilon_{i} \rho_{i}{\dot{\bf{G}}}+\beta_{gi}({\bf{u}}_g-{\bf{u}}_i)+\sum_{j=1}^2\beta_{ij}({\bf{u}}_j-{\bf{u}}_i)$\\
\\
Granular temperature equation of solid species $i$:\\\\
$\frac{3}{2}[\frac{\partial}{\partial t}(\varepsilon_{i}\rho_i \theta_{i})+\nabla \cdot(\varepsilon_{i}\rho_i \theta_{i} \mathbf{u}_{i})]=-{\bf{P}}_i: \nabla \mathbf{u}_{i}+\nabla \cdot \mathbf{q}_{i}-\gamma_{i}-3\beta_{gi}\theta_i$\\\\
\hline
\end{tabular*}
\begin{tablenotes}    % 添加命令
        \footnotesize               % 添加命令
        *The subscript $i$ means hydrodynamic characters of solid species $i$ and subscript $g$ means the property of gas.\\       %自己改注释 和表格对应
        *The pressure and shear stress tensor of gas phase are denoted as $p_g$ and ${\bm{\tau}}_g\equiv\mu_g(\nabla {\bf{u}}_g+\nabla{\bf{u}}_g^T)+(\lambda_g-\frac{2}{3}\mu_g)(\nabla\cdot{\bf{u}}_g){\bf{I}}$, respectively; The  gas-solid drag force coefficient of solid species $i$ is $\beta_{gi}$, the solid-solid drag force coefficient of species $i$ in term of particles interaction between species $i$ and species $j$ is $\beta_{ij}$\\  		  % 自己改注释 和表格对应
        *The energy dissipation for gas-solid interaction is $-\beta_{gi}\theta_i$.\\	  %自己改注释 和表格对应
        {{*Following most of the available studies in literature \citep{wang2020continuum}, we have neglected the effect of gas turbulence. The main reason is that the fluctuation velocity of particles and the volume fraction of particles are correspondingly comparative to the fluctuation velocity and the volume fraction of gas, whereas the density of particles are three orders of magnitude larger than that of gas. Therefore, the stress of particle is significantly larger than the Reynolds stress of gases.}}
 \end{tablenotes}
\end{center}
\end{table}

Considering that the solid constitutive relations (\ref{a26})-(\ref{a36}) are too complex to couple with {CFD simulation by the commercial software FLUNET, because we can only implement our model via UDF}, we first attempt to simplify these relations as follows:\\
(i) The second-order tensor that relates to the slip velocity of different species $({\bf{u}}_j-{\bf{u}}_i)({\bf{u}}_j-{\bf{u}}_i)$ is assumed to be a diagonally dominant matrix, as discussed in \cite{wang2010flow}, where it was shown that  the normal stress is much larger than the shear stress. In addition, the slip velocity $({\bf{u}}_j-{\bf{u}}_i)$ is assumed to be isotropic according to previous studies \citep{gidaspow1994multiphase}. Therefore, this second-order tensor is simplified as $({\bf{u}}_j-{\bf{u}}_i)({\bf{u}}_j-{\bf{u}}_i)\approx\frac{1}{3}({\bf{u}}_j-{\bf{u}}_i)^2\bf{I}$.\\
(ii) Referring to the previous studies \citep{mathiesen2000experimental,he2020unified,zhao2021multiscale}, the gradients of hydrodynamic variables of species $i$ and $j$ are assumed to be equivalent to the corresponding values of granular mixture, we then have  $\nabla\varepsilon_i=\nabla\varepsilon_j=\nabla\varepsilon$, $\nabla{\bf{u}}_i=\nabla{\bf{u}}_j=\nabla{\bf{u}}$ and $\nabla\theta_i=\nabla\theta_j=\nabla\theta$.\\
(iii) The first-order inequalities of hydrodynamic variables include
$\varepsilon_i-\varepsilon_j$, ${\bf{u}}_i-{\bf{u}}_j$, $\theta_i-\theta_j$, and $\nabla\varepsilon_i$, $\nabla{\bf{u}}_i$, $\nabla\theta_i$.  Higher order inequalities can be constructed from these first-order inequalities, but for the sake of simplifying constitutive relations,
all higher order terms, except for $({\bf{u}}_j-{\bf{u}}_i)^2$, have been disregarded.\\
(iv) According to the fundamental concept of kinetic theory, the constitutive relations at the $O(1)$ order are more crucial than those at the $O(K)$ order. Therefore, the $O(K)$ order contributions are neglected for the ease of coupling present constitutive relations into multifluid model simulations.

Given the above assumptions, the solid constitutive relations for the multifluid model in binary gas-solid flow can be simplified as follows:\\
The stress tensor for species $i$ is expressed as:
\begin{equation}
\begin{split}
&{\bf{P}}_i=p_{si}{\bf{I}}-2\mu_i\varepsilon_i\mathring{\overline{\overline{\nabla{\bf{u}}_i}}}-\lambda_i\varepsilon_i(\nabla\cdot{\bf{u}}_i){\bf{I}},
\end{split}
\label{a51}
\end{equation}
where the corresponding stress strain is denoted as $\mathring{\overline{\overline{\nabla{\bf{u}}_i}}}\equiv\frac{1}{2}(\nabla {\bf{u}}_i+\nabla {\bf{u}}_i^T)-\frac{1}{3}(\nabla\cdot{\bf{u}}_i){\bf{I}}$ , the normal pressure $p_{si}$, the shear viscosity $\mu_i$ and the bulk viscosity of species $i$ are:
\begin{equation}
\begin{split}
p_{si}=\varepsilon_i\rho_i\theta_i+\sum_{j=1}^{2}\bigg\{\frac{\pi}{3}(1+e_{ij})d_{ij}^3g_{ij}\frac{\varepsilon_i\varepsilon_j\rho_i\rho_j}{m_0}
\bigg[(\theta_i+\theta_j)+\frac{1}{3}({\bf{u}}_j-{\bf{u}}_i)^2\bigg]\bigg\},
\end{split}
\label{a52}
\end{equation}
\begin{equation}
\begin{split}
\mu_{i}=\sum_{j=1}^{2}\bigg\{\frac{1}{3}2^{-{\frac{3}{2}}}{\pi}^{-\frac{1}{2}}(\theta_i+\theta_j)^{\frac{1}{2}}(1+e_{ij})d_{ij}^4g_{ij}(\frac{\varepsilon_j\rho_i\rho_j}{m_0}) \exp(-|\overline{{\bf{g}}^{*}}|^2)[\frac{1}{4}H_8+\pi M_1]\bigg\},
\end{split}
\label{a53}
\end{equation}
\begin{equation}
\begin{split}
\lambda_i=\sum_{j=1}^2\bigg\{\frac{5}{18}2^{-{\frac{1}{2}}}{\pi}^{-\frac{1}{2}}(\theta_i+\theta_j)^{\frac{1}{2}}(1+e_{ij})d_{ij}^4g_{ij}(\frac{\varepsilon_j\rho_i\rho_j}{m_0}) \exp(-|\overline{{\bf{g}}^{*}}|^2)[\frac{1}{4}H_8+\pi M_1]\bigg\}.
\end{split}
\label{a54}
\end{equation}\\
The granular heat flux of species $i$ is
\begin{equation}
\begin{split}
&{\bf{q}}_{i}={\bf{q}}_{i1}+{\bf{q}}_{i2},\\
&{\bf{q}}_{i1}=-\sum_{j=1}^2\bigg\{\frac{2^{-\frac{7}{2}}}{3}\pi^{\frac{1}{2}}d_{ij}^4m_ig_{ij}n_in_j[2\eta_i](\theta_i+\theta_j)^{\frac{3}{2}}\exp(-\overline{{g^*}}^{2})
\bigg\{\pi^{-1}\bigg[-A^{-1}\bigg(\frac{5}{8}H_8+\frac{3}{2}\pi h_6
\bigg)\bigg]-\bigg[A^{-1}h_6\bigg]\bigg\}{\frac{\nabla\theta}{\theta_{i}\theta_{j}}}\bigg\},\\
&{\bf{q}}_{i2}=\sum_{j=1}^2\bigg\{\frac{\pi}{20}d_{ij}^3m_ig_{ij}n_in_j
\bigg[[\eta_i^2(\theta_i+\theta_j)-2\eta_i\theta_i]2(5+2\bar{g}^{*2})
+\frac{20\eta_i}{3}\theta_i\bigg]({\bf{u}}_j-{\bf{u}_i}).\\
\end{split}
\label{a55}
\end{equation}
Referring to the implementation into FLUENT (student version), the heat flux in the granular temperature equation is expressed by the Fourier's law as $\mathbf{q}_i=-k_i\nabla\theta_i$ in Fluent. It can be seen that ${\bf{q}}_{i1}$ can be easily implemented using the User-defined functions (UDF) by installing the specific Granular Conductivity, whereas ${\bf{q}}_{i2}$  is fundamentally inconsistent to the Fourier's law, therefore, $\nabla\cdot{\bf{q}}_{i2}$ is implemented as a Source Term of granular temperature equation.\\
The particle-particle drag coefficient is:
\begin{equation}
\begin{split}
\beta_{ij}=\sum_{j=1}^2\bigg\{{(1+e_{ij})d_{ij}^2}g_{ij}\frac{\varepsilon_i\varepsilon_j\rho_i\rho_j}{m_0}(\theta_i+\theta_j)^{\frac{1}{2}}\pi^{\frac{1}{2}}
\bigg[(1+\frac{1}{2\overline{{g^*}}^{2}})\exp(-\overline{{g^*}}^{2})+\sqrt{\pi}(\frac{1}{\overline{{g^*}}}+\overline{{g^*}}
-\frac{1}{4\overline{{g^*}}^{3}})erf(\overline{{g^*}})\bigg]2^{-\frac{1}{2}}\bigg\},
\end{split}
\label{a56}
\end{equation}
and the source term of energy dissipation rate is:
\begin{equation}
\begin{split}
\gamma_{i}=&-\sum_{j=1}^2\bigg\{
3\pi^{-\frac{1}{2}}g_{ij}\varepsilon_i\varepsilon_j\rho_i\frac{d_{ij}^2}{d_{j}^3}
\bigg[\bigg(\eta_i^2(\theta_i+\theta_j)-2\eta_i\theta_i\bigg){[2(\theta_i+\theta_j)]}^{\frac{1}{2}}\mathcal{L}_3
+2\eta_i\bigg[\frac{1}{2(\theta_i+\theta_j)}\bigg]^{\frac{1}{2}}\bigg((\theta_i{\bf{u}}_j+\theta_j{\bf{u}}_i)\cdot({\bf{u}}_j-{\bf{u}}_i)\bigg)\mathcal{L}_2\bigg]\\
&\quad\quad+6g_{ij}\varepsilon_i\varepsilon_j\rho_i\bigg(\frac{d_{ij}}{d_j}\bigg)^3
\bigg[\eta_i^2(\theta_i+\theta_j)-2\eta_i\theta_i\bigg](-\nabla\cdot{\bf{u}}_i)\bigg\}.\\
\end{split}
\label{a57}
\end{equation}
And the specific expressions have been provided for the coefficient functions of dimensionless slip velocity, including $H_8$, $M_1$, $h_6$, $\mathcal{L}_2$, and $\mathcal{L}_3$ in the preceding section.

In those constitutive relations, the radial distribution function $g_{ij}$ proposed by \cite{lun1984kinetic} is used:
\begin{equation}
\begin{split}
g_{ij}=\frac{d_{ii}g_{jj}+d_{jj}g_{ii}}{d_i+d_j}, \ \ g_{ii}=\bigg[1-\big(\frac{\varepsilon_s}{\varepsilon_{s,max}}\big)^\frac{1}{3}\bigg]^{-1}+\frac{1}{2}d_i\sum_{j=1}^2\frac{\varepsilon_j}{d_j}.
\end{split}
\label{a58}
\end{equation}

As we all known, the frictional stress model is critical when the total solid concentration is higher than a specific value. An $ad \ hoc$ extension of the frictional viscosity $\mu_{i,f}$ \citep{schaeffer1987instability} and the frictional pressure $p_{si,f}$ \citep{johnson1987frictional} for monodisperse flow to binary gas-solid flow are made:
\begin{equation}
\begin{split}
\mu_{i,f} = \frac{p_{si,f}sin\phi}{2\sqrt{I_{2D}}},
\end{split}
\label{a63}
\end{equation}
\begin{equation}
\begin{split}
p_{si,f} = \varepsilon_{i}Fr\frac{(\varepsilon_{s}-\varepsilon_{s,min})^n}{(\varepsilon_{s,max}-\varepsilon_{s})^p}.
\end{split}
\label{a64}
\end{equation}
where $\phi$ is the angle of internal friction, $I_{2D}$ is the second invariant of the deviatoric stress tensor, $\varepsilon_{s,max}$ is the solid packing limit, $\varepsilon_{s,min}=0.5$ is the minimum of solid concentrated required to consider the effect of particle friction, $\varepsilon_{s}\equiv\sum_{i=1}^2\varepsilon_i$ is the volume fraction of solid mixture, and  the related parameters are $Fr=0.05$, the coefficient $n=2$ and $p=5$.

The gas-solid drag force is also crucial to a correct prediction of particle behavior and the Gidaspow's model \citep{gidaspow1994multiphase} is used:
\begin{equation}
\beta_{gi} =
\left\{
\begin{aligned}
&\frac{3}{4}C_{d,i}\frac{\varepsilon_i\varepsilon_g\rho_g|{\bf{u}}_i-{\bf{u}}_g|}{d_i}\varepsilon_g^{-2.65} &\textrm{$\varepsilon_g\geq0.8$}\\
&150\frac{\varepsilon_i(1-\varepsilon_g)\mu_g}{\varepsilon_gd_i^2}+1.75\frac{\varepsilon_i\rho_g|{\bf{u}}_g-{\bf{u}}_i|}{d_i} &\textrm{$\varepsilon_g<0.8$}
\end{aligned}
\right.
\label{a60}
\end{equation}
where the drag coefficient is:
\begin{equation}
C_{d,i} =
\left\{
\begin{aligned}
&\frac{24}{Re_i}[1+0.15(Re_i)^{0.687}]
&\textrm{$Re_i\leq1000$}\\
&0.44
&\textrm{$Re_i>1000$},
\end{aligned}
\right.
\label{a60}
\end{equation}
and the Reynolds number is
\begin{equation}
\begin{split}
&Re_i=\frac{d_i\rho_g\varepsilon_g|{\bf{u}}_g-{\bf{u}}_i|}{\mu_g}.\\
\end{split}
\label{a62}
\end{equation}

\subsection{Simulation layout}
Fluidization experiments on binary Geldart group B mixtures were conducted in a Plexiglas column by \cite{joseph2007experimental}, their results are utilized to validate present multifluid model.
The jetsam is glass particles with a diameter of 116 $\mu m$, while the flotsam is polystyrene particles with a diameter of 275 $\mu m$.
The simulations are performed in two-dimensional domain using commercial software FLUENT, with a bed height of 0.8m and bed diameter of 0.184m.
The above governing equations are solved using the finite-volume method, where the phase-coupled SIMPLE algorithm  is used to solve the pressure-velocity coupling problem. The momentum equations, volume fraction equations and granular temperature equations are discretized using the QUICK scheme. The wall boundary conditions for gas and solid phases are respectively no slip wall condition and free slip wall condition. {{The grid dependency of this binary fluidized bed has been investigated in the previous work \citep{zhong2012cfd}, who study has concluded that the grid number ($37\times80$ ) is sufficient to provide a grid-size-independent results. Thus, we also set the same grid number }} and the time step is 0.001s. The physical characteristics of gas and solid particles, along with other parameters used in numerical simulations, are outlined in Tab.\ref{tabA2}.
\begin{table}[h]
\caption{Physical properties and simulation parameters used in the simulations.}
\label{tabA2}
\begin{center}
\begin{tabular*}{0.9\hsize}{@{}@{\extracolsep{\fill}}llllll@{}}
 \hline
Particle properties &jetsam &flotsam  \\ \hline
Particle diameter ($\mu m$), $d_i$ &116 &275\\
Particle density ($kg/m^3$), $\rho_i$ &2476 &1064 \\
Minimum fluidization velocity(m/s), $u_{mf}$	&0.018	&0.04 \\
Restitution coefficient (-), e	&0.99	&0.99	 \\
Mass fraction  (-), mof	&0.69	&0.31 \\
Volume fraction  (-), vof	&0.50	&0.50\\
\hline
\multicolumn{3}{l}{Gas properties} \\ \hline
Superficial gas velocity (m/s), $U_g$	&0.078 & 0.056 \\
Gas viscosity ($Pa\cdot s$), $\mu_g$	&\multicolumn{2}{c}{$1.83\times10^{-5}$}  \\
Gas density ($kg/m^3$), $\rho_g$	&\multicolumn{2}{c}{1.0}  \\
\hline
\multicolumn{3}{l}{Others equipment parameters} \\ \hline
Column height (m), H	&\multicolumn{2}{c}{0.4}\\
Column diameter (m), D	&\multicolumn{2}{c}{0.184}  \\
Initial packing height (m), $H_i$	&\multicolumn{2}{c}{0.22}   \\
The bed expansion height (m),$H_{fc}$	&\multicolumn{2}{c}{0.232}	 \\
Packed bed void fraction (-), $\varepsilon_{g}^*$	&\multicolumn{2}{c}{0.41} \\
\hline
\end{tabular*}
\end{center}
\end{table}

\subsection{Simulation results}
\begin{figure}[H]
\centerline{\includegraphics[width=0.55\textwidth]{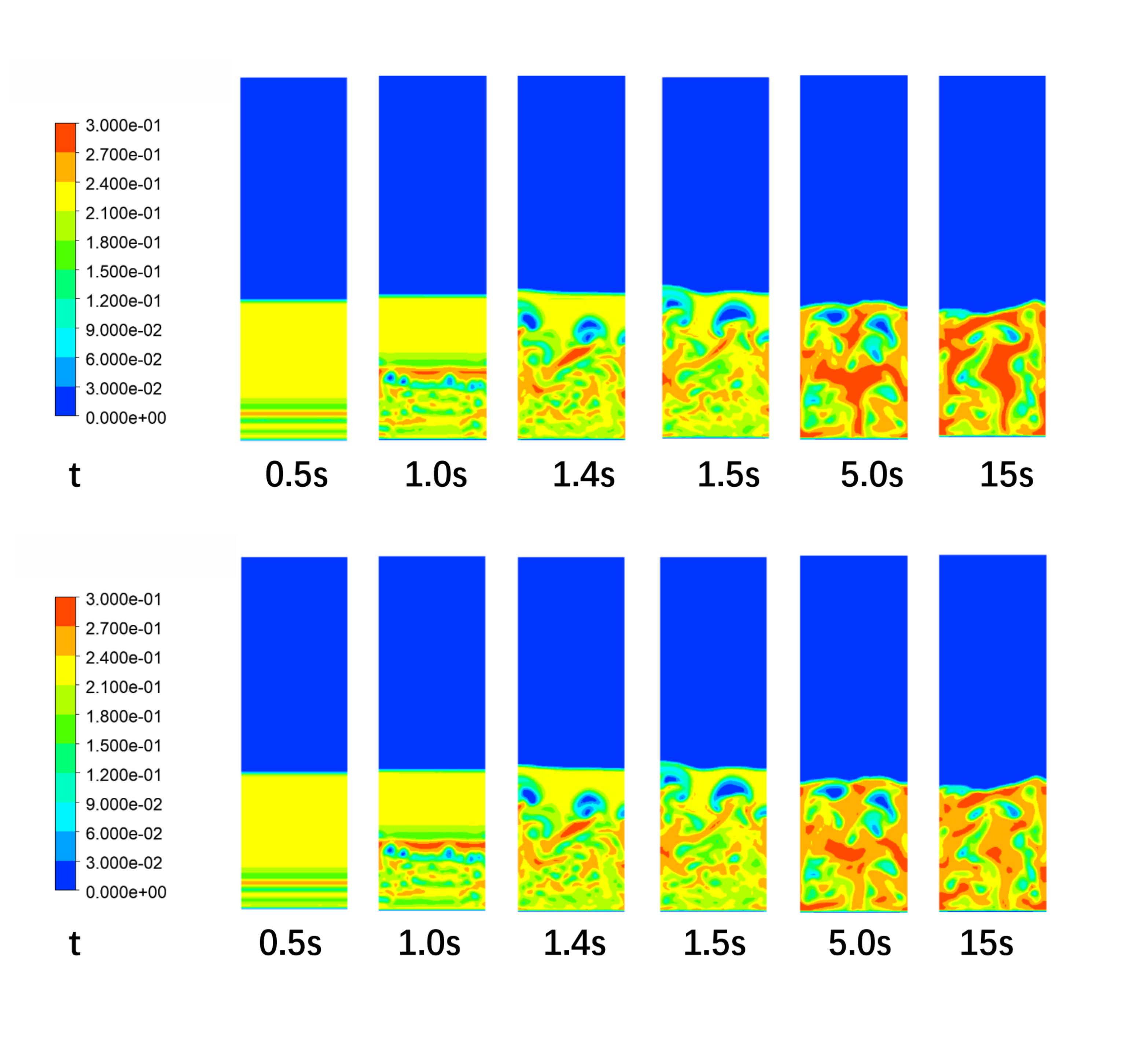}}
\caption{Snapshots of solid volume fraction, top:  jetsam; bottom: flotsam.}\label{figb_2}
\end{figure}

The contours plotted in Figure \ref{figb_2} display the snapshots of solid volume fraction at various simulation times for the binary mixture {{at $U_g=$0.078m/s}}.
At the beginning of fluidization, one-dimensional traveling wave is formed due to strong particle-particle and gas-particle interactions (t=0.5 s), which is then further destabilized to form small bubbles and voids (t=1.0 s), those bubbles and voids rise to the top of the bed under significant coalescence as shown in the snapshots of 1.4 s and 1.5 s, thus leading to the formation of larger bubbles. After few seconds, the bed operates at a standard bubbling fluidization regime with vigorous bubble motions. Moreover, it can be seen that the instability is firstly triggered at the bottom part of the bed and then gradually propagates to the entire bed. All those phenomena are qualitatively correct.

\begin{figure}[H]
\centerline{\includegraphics[width=0.6\textwidth]{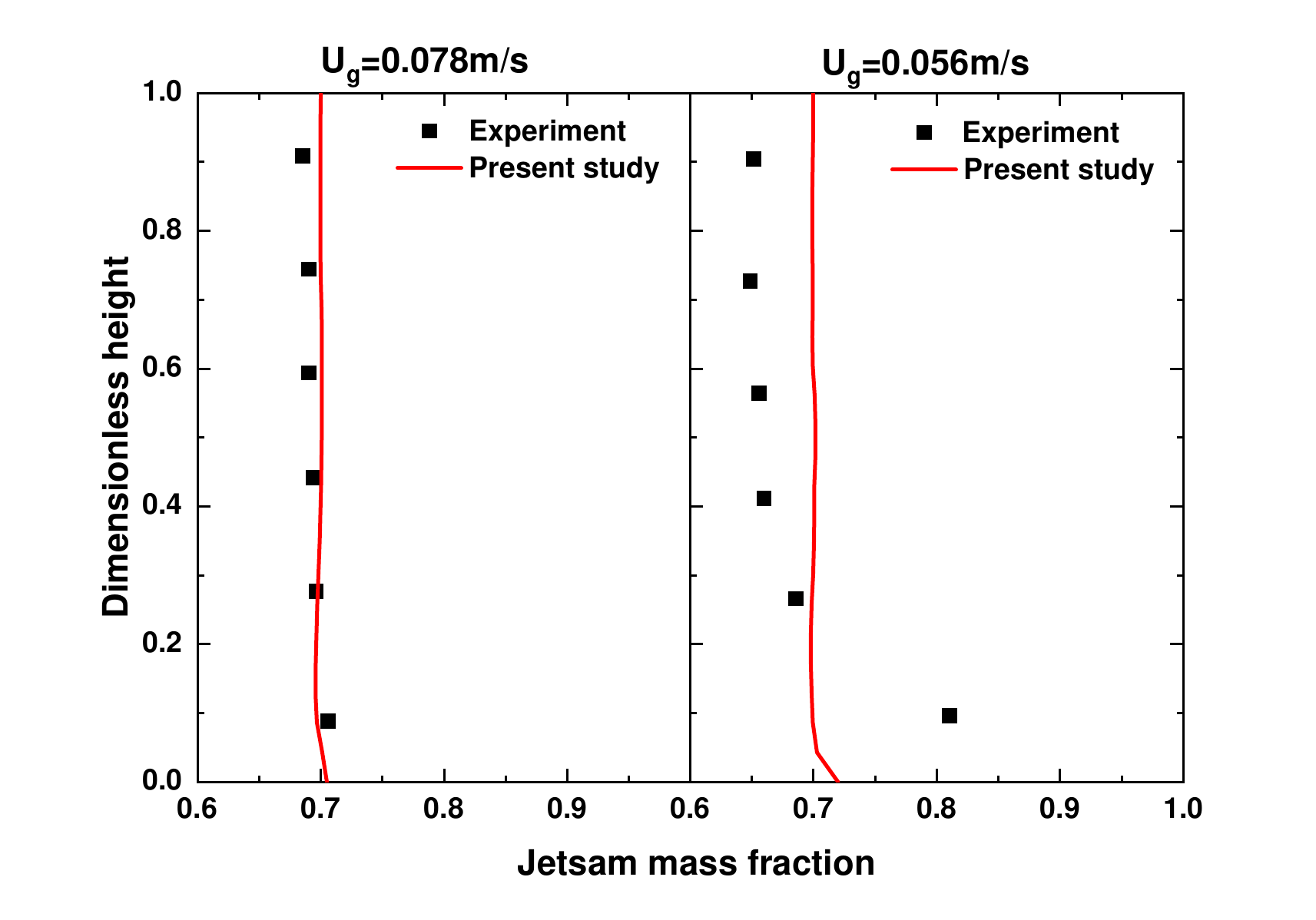}}
\caption{Axial  distributions of jetsam mass fraction at different superficial gas velocities $U_g$.}\label{figb_3}
\end{figure}

\begin{figure}[H]
\centerline{\includegraphics[width=0.45\textwidth]{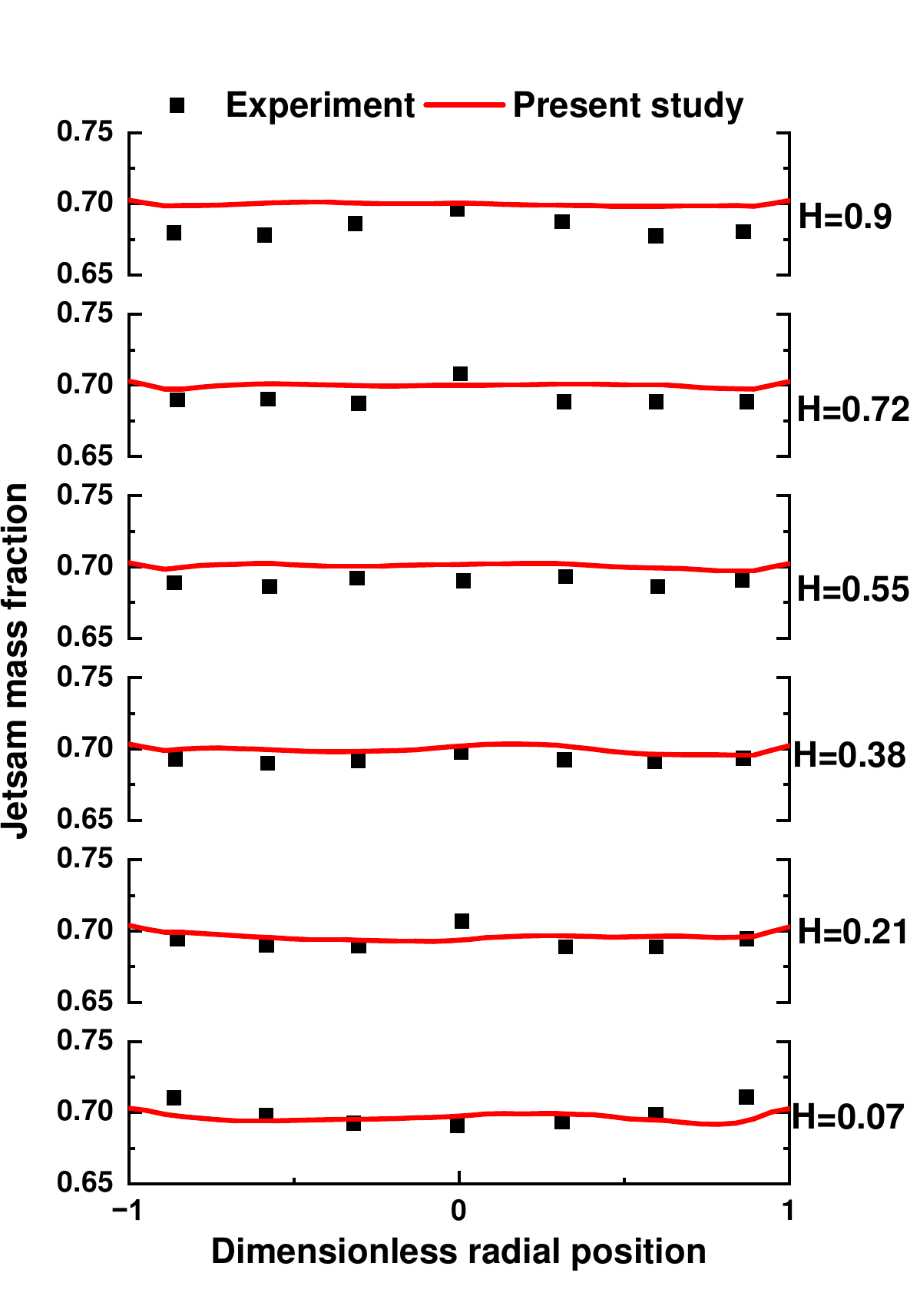}}
\caption{Radial distributions of jetsam mass fraction at the different bed height for $U_g=0.078m/s$.}\label{figb_4}
\end{figure}

Fig.(\ref{figb_3}) and Fig.(\ref{figb_4}) display the axial and radial distribution of the jetsam mass fraction, defined as the proportional mass fraction of jetsam in the overall solid mass fraction. it is easy to observe that (i) the multifluid predictions are in a good agreement with the experimental results of \cite{joseph2007experimental}, thus preliminarily but quantitatively validating the newly developed multifluid model; (ii) the two cases studied here correspond to well mixing conditions, where strong bubble motions enable sufficient mixing of binary species at higher superficial gas velocity{{; (iii) the possible reason for the simulation results of a low superficial velocity ($U_g=$0.056 m/s) case being not as good as that of a high ones ($U_g=$0.078 m/s) can be analysed as follows: the kinetic theory of polydisperse granular flow is only one of the critical inputs of multifluid model, other model selections such as gas-solid drag model and the frictional stress model can also have a significant effect on the simulation results. Furthermore, in addition to the model selection, the selection of input parameters will also affect the simulation result, such as the restitution coefficients. Therefore, it is not easy to find the exact source(s) of discrepancy, we prefer to believe that the exact source(s) of simulation discrepancy is/are unidentified.}}

\section{Conclusion}
Single-parameter Chapman-Enskog method with a new mathematical technique was used to develop a refined kinetic theory of polydisperse granular flows. This allowed the determination of constitutive relations analytically, including the solid stress tensor, the solid-solid drag force, the heat flux, and the energy dissipation rate, which are necessary for closing the multifluid model in polydisperse gas-solid flow.
{{This refined KTGF has been directly validated using DEM data of one-dimensional shear flow.}}
Then, the theory was coupled with multifluid model to simulate the hydrodynamics of binary gas-solid flow in a bubbling fluidized bed. The adequate mixing phenomenon of binary solid mixture was precisely described, thus providing a preliminary CFD validation of the polydisperse kinetic theory. Our future works will focus on the study about how to properly consider the effect of inter-species collision on the perturbation function, which appears to be the major deficiency of present theory. Furthermore, a particle-wall boundary condition model for polydisperse flow is not available yet. Both of those model (or theory) development will further improve the accuracy of multifluid model. {{In the future, we plan to couple the complete and refined polydisperse kinetic model with opensource platform OpenFOAM for CFD simulations.}}

%\section*{Declaration of Competing Interest}
%There is no conflict of interests.

\section*{Acknowledgement}
This study is financially supported by the Strategic Priority Research Program of the Chinese Academy of Sciences (XDA29040200), the Young Elite Scientists Sponsorship Program by CAST (2022QNRC001), the National Natural Science Foundation of China (22378399), and the National Key R$\&$D Program of China (2021YFB1715500).

\section*{Reference}

%%\nocite{*}
\bibliographystyle{elsarticle-harv}
\bibliography{Reference}

\begin{thebibliography}{50}
\expandafter\ifx\csname natexlab\endcsname\relax\def\natexlab#1{#1}\fi
\expandafter\ifx\csname url\endcsname\relax
  \def\url#1{\texttt{#1}}\fi
\expandafter\ifx\csname urlprefix\endcsname\relax\def\urlprefix{URL }\fi

\bibitem[{Beetstra et~al.(2007)Beetstra, Hoef, and Kuipers}]{Beetstra2007Drag}
Beetstra, R., Hoef, M. A. V.~D., Kuipers, J. A.~M., 2007. Drag force of
  intermediate {Reynolds} number flow past mono- and bidisperse arrays of
  spheres. AIChE Journal 53~(2), 489--501.

\bibitem[{Berger and Hrenya(2014)}]{berger2014challenges}
Berger, K.~J., Hrenya, C.~M., 2014. Challenges of {DEM: II. Wide} particle size
  distributions. Powder Technology 264, 627--633.

\bibitem[{Chao et~al.(2011)Chao, Wang, Jakobsen, Fernandino, and
  Jakobsen}]{chao2011derivation}
Chao, Z., Wang, Y., Jakobsen, J.~P., Fernandino, M., Jakobsen, H.~A., 2011.
  Derivation and validation of a binary multi-fluid {Eulerian} model for
  fluidized beds. Chemical Engineering Science 66~(16), 3605--3616.

\bibitem[{Chapman and Cowling(1970)}]{chapman1970mathematical}
Chapman, S., Cowling, T., 1970. The mathematical theory of non uniform gases,
  cambridge mathematical library.

\bibitem[{Galvin(2007)}]{galvin2007hydrodynamic}
Galvin, J.~E., 2007. On the hydrodynamic description of binary mixtures of
  rapid granular flows and gas-fluidized beds. Ph.D. thesis, University of
  Colorado at Boulder.

\bibitem[{Garz{\'o}(2019)}]{garzo2019navier}
Garz{\'o}, V., 2019. {Navier--Stokes Transport Coefficients for Multicomponent
  Granular Gases. I. Theoretical Results}. In: Granular Gaseous Flows.
  Springer, pp. 177--216.

\bibitem[{Garz{\'o}(2021)}]{V2021Comment}
Garz{\'o}, V., 2021. Comment on "kinetic theory models for granular mixtures
  with unequal granular temperature: Hydrodynamic velocity". Physics of Fluids
  33, 089101.

\bibitem[{Garz{\'o} et~al.(2007)Garz{\'o}, Dufty, and Hrenya}]{garzo2007enskog}
Garz{\'o}, V., Dufty, J.~W., Hrenya, C.~M., 2007. Enskog theory for
  polydisperse granular mixtures. {I. Navier-Stokes} order transport. Physical
  Review E 76~(3), 031303.

\bibitem[{Gidaspow(1994)}]{gidaspow1994multiphase}
Gidaspow, D., 1994. Multiphase flow and fluidization: continuum and kinetic
  theory descriptions. Academic press.

\bibitem[{Grace and Sun(1991)}]{grace1991influence}
Grace, J., Sun, G., 1991. Influence of particle size distribution on the
  performance of fluidized bed reactors. The Canadian Journal of Chemical
  Engineering 69~(5), 1126--1134.

\bibitem[{He et~al.(2020)He, Zhao, and Wang}]{he2020unified}
He, M., Zhao, B., Wang, J., 2020. {A unified EMMS-based constitutive law for
  heterogeneous gas-solid flow in CFB risers}. Chemical Engineering Science
  225, 115797.

\bibitem[{He et~al.(2009)He, Deen, Annaland, and Kuipers}]{he2009gas}
He, Y., Deen, N., Annaland, M. v.~S., Kuipers, J., 2009. Gas-solid turbulent
  flow in a circulating fluidized bed riser: numerical study of binary particle
  systems. Industrial \& Engineering Chemistry Research 48~(17), 8098--8108.

\bibitem[{Hrenya(2011)}]{hrenya2011kinetic}
Hrenya, C.~M., 2011. Kinetic theory for granular materials: polydispersity. In:
  Computational gas-solids flows and reacting systems: Theory, methods and
  practice. IGI Global, pp. 102--127.

\bibitem[{Huilin et~al.(2001)Huilin, Gidaspow, and Manger}]{huilin2001kinetic}
Huilin, L., Gidaspow, D., Manger, E., 2001. Kinetic theory of fluidized binary
  granular mixtures. Physical Review E 64~(6), 061301.

\bibitem[{Huilin et~al.(2003)Huilin, Yurong, and
  Gidaspow}]{huilin2003hydrodynamic}
Huilin, L., Yurong, H., Gidaspow, D., 2003. Hydrodynamic modelling of binary
  mixture in a gas bubbling fluidized bed using the kinetic theory of granular
  flow. Chemical Engineering Science 58~(7), 1197--1205.

\bibitem[{Iddir(2004)}]{Iddir2004modeling}
Iddir, H., 2004. Modeling of the multiphase mixture of particles using the
  kinetic theory approach. Ph.D. thesis, Illinois Institute of Technology.

\bibitem[{Iddir and Arastoopour(2005)}]{iddir2005modeling}
Iddir, H., Arastoopour, H., 2005. Modeling of multitype particle flow using the
  kinetic theory approach. AIChE Journal 51~(6), 1620--1632.

\bibitem[{Iddir et~al.(2005)Iddir, Arastoopour, and Hrenya}]{iddir2005analysis}
Iddir, H., Arastoopour, H., Hrenya, C.~M., 2005. Analysis of binary and ternary
  granular mixtures behavior using the kinetic theory approach. Powder
  Technology 151~(1-3), 117--125.

\bibitem[{Jenkins and Mancini(1989)}]{jenkins1989kinetic}
Jenkins, J., Mancini, F., 1989. Kinetic theory for binary mixtures of smooth,
  nearly elastic spheres. Physics of Fluids A: Fluid Dynamics 1~(12),
  2050--2057.

\bibitem[{Johnson and Jackson(1987)}]{johnson1987frictional}
Johnson, P.~C., Jackson, R., 1987. Frictional--collisional constitutive
  relations for granular materials, with application to plane shearing. Journal
  of fluid Mechanics 176, 67--93.

\bibitem[{Joseph et~al.(2007)Joseph, Leboreiro, Hrenya, and
  Stevens}]{joseph2007experimental}
Joseph, G.~G., Leboreiro, J., Hrenya, C.~M., Stevens, A.~R., 2007. Experimental
  segregation profiles in bubbling gas-fluidized beds. AIChE journal 53~(11),
  2804--2813.

\bibitem[{Lan et~al.(2021)Lan, Xu, Zhao, Zou, Wang, and Zhu}]{lan2021scale}
Lan, B., Xu, J., Zhao, P., Zou, Z., Wang, J., Zhu, Q., 2021. Scale-up effect of
  residence time distribution of polydisperse particles in continuously
  operated multiple-chamber fluidized beds. Chemical Engineering Science 244,
  116809.

\bibitem[{Lan et~al.(2020)Lan, Xu, Zhao, Zou, Zhu, and Wang}]{lan2020long}
Lan, B., Xu, J., Zhao, P., Zou, Z., Zhu, Q., Wang, J., 2020. Long-time
  coarse-grained {CFD-DEM} simulation of residence time distribution of
  polydisperse particles in a continuously operated multiple-chamber fluidized
  bed. Chemical Engineering Science 219, 115599.

\bibitem[{Lathouwers and Bellan(2000)}]{lathouwers2000modeling}
Lathouwers, D., Bellan, J., 2000. Modeling and simulation of bubbling fluidized
  beds containing particle mixtures. Proceedings of the Combustion Institute
  28~(2), 2297--2304.

\bibitem[{Lun et~al.(1984)Lun, Savage, Jeffrey, and Chepurniy}]{lun1984kinetic}
Lun, C., Savage, S.~B., Jeffrey, D., Chepurniy, N., 1984. Kinetic theories for
  granular flow: inelastic particles in couette flow and slightly inelastic
  particles in a general flowfield. Journal of Fluid Mechanics 140, 223--256.

\bibitem[{Mathiesen et~al.(2000{\natexlab{a}})Mathiesen, Solberg, and
  Hjertager}]{mathiesen2000experimental}
Mathiesen, V., Solberg, T., Hjertager, B.~H., 2000{\natexlab{a}}. An
  experimental and computational study of multiphase flow behavior in a
  circulating fluidized bed. International Journal of Multiphase Flow 26~(3),
  387--419.

\bibitem[{Mathiesen et~al.(2000{\natexlab{b}})Mathiesen, Solberg, and
  Hjertager}]{mathiesen2000predictions}
Mathiesen, V., Solberg, T., Hjertager, B.~H., 2000{\natexlab{b}}. {Predictions
  of gas/particle flow with an Eulerian model including a realistic particle
  size distribution}. Powder Technology 112~(1-2), 34--45.

\bibitem[{Qin et~al.(2019)Qin, Zhou, and Wang}]{qin2019emms}
Qin, Z., Zhou, Q., Wang, J., 2019. An {EMMS} drag model for coarse grid
  simulation of polydisperse gas--solid flow in circulating fluidized bed
  risers. Chemical Engineering Science 207, 358--378.

\bibitem[{Rosato et~al.(1987)Rosato, Strandburg, Prinz, and
  Swendsen}]{rosato1987brazil}
Rosato, A., Strandburg, K.~J., Prinz, F., Swendsen, R.~H., 1987. Why the brazil
  nuts are on top: Size segregation of particulate matter by shaking. Physical
  review letters 58~(10), 1038.

\bibitem[{Schaeffer(1987)}]{schaeffer1987instability}
Schaeffer, D.~G., 1987. Instability in the evolution equations describing
  incompressible granular flow. Journal of differential equations 66~(1),
  19--50.

\bibitem[{Sela and Goldhirsch(1998)}]{sela1998hydrodynamic}
Sela, N., Goldhirsch, I., 1998. Hydrodynamic equations for rapid flows of
  smooth inelastic spheres, to {Burnett} order. Journal of Fluid Mechanics 361,
  41--74.

\bibitem[{Serero et~al.(2006)Serero, Goldhirsch, Noskowicz, and
  Tan}]{serero2006hydrodynamics}
Serero, D., Goldhirsch, I., Noskowicz, S., Tan, M.-L., 2006. Hydrodynamics of
  granular gases and granular gas mixtures. Journal of Fluid Mechanics 554,
  237--258.

\bibitem[{Shi et~al.(2022)Shi, He, Zhang, Zhao, and Wang}]{shi2022critical}
Shi, K., He, M., Zhang, L., Zhao, B., Wang, J., 2022. {Critical comparison of
  polydisperse kinetic theories using bidisperse DEM data}. Chemical
  Engineering Science 263, 118062.

\bibitem[{Shu et~al.(2014)Shu, Wang, Fan, and Li}]{shu2014multifluid}
Shu, Z., Wang, J., Fan, C., Li, S., 2014. Multifluid modeling of mixing and
  segregation of binary gas--solid flow in a downer reactor for coal pyrolysis.
  Industrial \& Engineering Chemistry Research 53~(23), 9915--9924.

\bibitem[{Solsvik and Manger(2021{\natexlab{a}})}]{solsvik2021kinetica}
Solsvik, J., Manger, E., 2021{\natexlab{a}}. {Kinetic theory models for
  granular mixtures with unequal granular temperature: Derivation of analytical
  constitutive equations}. Powder Technology 385, 580--597.

\bibitem[{Solsvik and Manger(2021{\natexlab{b}})}]{solsvik2021kineticb}
Solsvik, J., Manger, E., 2021{\natexlab{b}}. Kinetic theory models for granular
  mixtures with unequal granular temperature: Hydrodynamic velocity. Physics of
  Fluids 33~(4), 043321.

\bibitem[{Sun and Grace(1990)}]{sun1990effect}
Sun, G., Grace, J.~R., 1990. The effect of particle size distribution on the
  performance of a catalytic fluidized bed reactor. Chemical Engineering
  Science 45~(8), 2187--2194.

\bibitem[{Sundaresan(2001)}]{sundaresan2001some}
Sundaresan, S., 2001. Some outstanding questions in handling of cohesionless
  particles. Powder Technology 115~(1), 2--7.

\bibitem[{van Sint~Annaland et~al.(2009{\natexlab{a}})van Sint~Annaland,
  Bokkers, Goldschmidt, Olaofe, van~der Hoef, and
  Kuipers}]{van2009developmentI}
van Sint~Annaland, M., Bokkers, G., Goldschmidt, M., Olaofe, O., van~der Hoef,
  M.~A., Kuipers, J., 2009{\natexlab{a}}. Development of a multi-fluid model
  for polydisperse dense gas--solid fluidised beds, {Part I: Model} derivation
  and numerical implementation. Chemical Engineering Science 64~(20),
  4222--4236.

\bibitem[{van Sint~Annaland et~al.(2009{\natexlab{b}})van Sint~Annaland,
  Bokkers, Goldschmidt, Olaofe, van~der Hoef, and
  Kuipers}]{van2009developmentII}
van Sint~Annaland, M., Bokkers, G., Goldschmidt, M., Olaofe, O., van~der Hoef,
  M.~A., Kuipers, J., 2009{\natexlab{b}}. Development of a multi-fluid model
  for polydisperse dense gas--solid fluidised beds, {Part II: Segregation} in
  binary particle mixtures. Chemical Engineering Science 64~(20), 4237--4246.

\bibitem[{Wang(2010)}]{wang2010flow}
Wang, J., 2010. {Flow structures inside a large-scale turbulent fluidized bed
  of FCC particles: Eulerian simulation with an EMMS-based sub-grid scale
  model}. Particuology 8~(2), 176--185.

\bibitem[{Wang(2020)}]{wang2020continuum}
Wang, J., 2020. Continuum theory for dense gas-solid flow: A state-of-the-art
  review. Chemical Engineering Science 215, 115428.

\bibitem[{Willits and Arnarson(1999)}]{willits1999kinetic}
Willits, J.~T., Arnarson, B., 1999. Kinetic theory of a binary mixture of
  nearly elastic disks. Physics of Fluids 11~(10), 3116--3122.

\bibitem[{Zhang et~al.(2020)Zhang, Jia, Xu, Wang, Duan, Ge, and
  Zhao}]{zhang2020cfd}
Zhang, Y., Jia, Y., Xu, J., Wang, J., Duan, C., Ge, W., Zhao, Y., 2020. {CFD}
  intensification of coal beneficiation process in gas-solid fluidized beds.
  Chemical Engineering and Processing-Process Intensification 148, 107825.

\bibitem[{Zhao et~al.(2021)Zhao, He, and Wang}]{zhao2021multiscale}
Zhao, B., He, M., Wang, J., 2021. Multiscale kinetic theory for heterogeneous
  granular and gas-solid flows. Chemical Engineering Science 232, 116346.

\bibitem[{Zhao and Wang(2018)}]{zhao2018unification}
Zhao, B., Wang, J., 2018. Unification of particle velocity distribution
  functions in gas-solid flow. Chemical Engineering Science 177, 333--339.

\bibitem[{Zhao and Wang(2020)}]{zhao2020note}
Zhao, B., Wang, J., 2020. A note on the kinetic theory of polydisperse granular
  flow. Chemical Engineering Science 223, 115730.

\bibitem[{Zhao and Wang(2021)}]{zhao2021kinetic}
Zhao, B., Wang, J., 2021. Kinetic theory of polydisperse gas--solid flow:
  {Navier--Stokes} transport coefficients. Physics of Fluids 33~(10), 103322.

\bibitem[{Zhong et~al.(2012)Zhong, Gao, Xu, and Lan}]{zhong2012cfd}
Zhong, H., Gao, J., Xu, C., Lan, X., 2012. Cfd modeling the hydrodynamics of
  binary particle mixtures in bubbling fluidized beds: Effect of wall boundary
  condition. Powder Technology 230, 232--240.

\bibitem[{Zhou and Wang(2015)}]{zhou2015cfd}
Zhou, Q., Wang, J., 2015. {CFD} study of mixing and segregation in {CFB}
  risers: extension of {EMMS} drag model to binary gas--solid flow. Chemical
  Engineering Science 122, 637--651.

\end{thebibliography}
\end{spacing}
\end{document}